\documentclass[aps,pre,reprint,superscriptaddress,amsmath,amssymb]{revtex4-2}

\usepackage{graphicx}
\usepackage{txfonts}
\usepackage[hypertexnames=false]{hyperref}
\usepackage{braket}
\usepackage{bm}
\usepackage{enumitem}

\begin{document}

\title{Power-law correlation in the homogeneous disordered state of anisotropically self-propelled systems}
\author{Kyosuke Adachi}
\affiliation{RIKEN Interdisciplinary Theoretical and Mathematical Sciences Program (iTHEMS), 2-1 Hirosawa, Wako 351-0198, Japan}
\affiliation{Nonequilibrium Physics of Living Matter RIKEN Hakubi Research Team, RIKEN Center for Biosystems Dynamics Research (BDR), 2-2-3 Minatojima-minamimachi, Chuo-ku, Kobe 650-0047, Japan}
\author{Hiroyoshi Nakano}
\affiliation{Institute for Solid State Physics, University of Tokyo, 5-1-5, Kashiwanoha, Kashiwa 277-8581, Japan}

\date{\today}

\begin{abstract}
Self-propelled particles display unique collective phenomena, due to the intrinsic coupling of density and polarity.
For instance, the giant number fluctuation appears in the orientationally ordered state, and the motility-induced phase separation appears in systems with repulsion.
Effects of strong noise typically lead to a homogeneous disordered state, in which the coupling of density and polarity can still play a significant role.
Here, we study universal properties of the homogeneous disordered state in two-dimensional systems with uniaxially anisotropic self-propulsion.
Using hydrodynamic arguments, we propose that the density correlation and polarity correlation generically exhibit power-law decay with distinct exponents ($-2$ and $-4$, respectively) through the coupling of density and polarity.
Simulations of self-propelled lattice gas models indeed show the predicted power-law correlations, regardless of whether the interaction type is repulsion or alignment.
Further, by mapping the model to a two-component boson system and employing non-Hermitian perturbation theory, we obtain the analytical expression for the structure factors, the Fourier transform of the correlation functions.
This reveals that even the first order of the interaction strength induces the power-law correlations.
\end{abstract}

\maketitle

\section{Introduction}

Active matter, a crowd of self-propelled entities, displays a variety of collective behaviors~\cite{Marchetti2013,Needleman2017,Gompper2020}.
Using minimal particle models of self-propelled elements~\cite{Vicsek1995} and hydrodynamic arguments~\cite{Toner1995,Toner2012}, extensive studies have been performed to elucidate the fundamental mechanisms and universal properties of collective phenomena in active matter systems~\cite{Chate2020}.
An intrinsic feature of active matter is the interplay of density and polarity, which emerges from the polar self-propelled motion of each particle.
The coupling of density and polarity produces distinctive fluctuation properties and spontaneous inhomogeneity, such as the giant number fluctuation in the orientationally ordered state~\cite{Chate2020}, motility-induced phase separation (MIPS)~\cite{Cates2015}, and (micro)phase separation in flocking~\cite{Solon2013,Solon2015a,Solon2015b} or active nematics~\cite{Ngo2014}.
Naturally, the density-polarity coupling is expected to play a significant role even in the homogeneous disordered state, which appears when noise predominates over the effects of interactions.

According to the studies on externally driven many-body systems~\cite{Garrido1990,Dorfman1994,Schmittmann1995,Schmittmann1998}, violation of the detailed balance induces the long-range density correlation with power-law decay, as long as spatial anisotropy exists in dynamics.
Recently, based on an analogy between externally driven force and anisotropic self-propulsion, we have proposed example models of active matter that show the same type of long-range density correlation.
Specifically, we have focused on the homogeneous disordered state of uniaxially self-propelled particles with repulsive interactions~\cite{Adachi2022,Nakano2024}.
Our results raise fundamental questions: (i) whether the power-law correlation is universal and independent of the interaction type (e.g., repulsion or alignment), and (ii) whether the anisotropic self-propulsion can lead to unique properties that have no counterparts in externally driven systems.
Broadly, the effects of spatial anisotropy on collective behaviors of active matter have been studied for the flocking transition~\cite{Brambati2022,Solon2022,Chatterjee2022}, active nematics~\cite{Mishra2014}, and MIPS~\cite{Broker2023,Othman2024}.

Stochastic many-body systems, including active matter models, are generically mapped to and can be analyzed as non-Hermitian quantum systems~\cite{Doi1976,Peliti1985,Tauber2014}.
A prototypical example is the correspondence between the asymmetric simple exclusion process and the XXZ quantum spin chain~\cite{Gwa1992,Sandow1994,Kim1995,Essler1996}.
More recently, extensive studies on non-Hermitian systems~\cite{Ashida2020} have facilitated mutual interactions between the classical and quantum domains, as exemplified by topologically protected edge modes~\cite{Murugan2017,Dasbiswas2018}.
Given such developments, mapping active matter models to quantum systems not only imports analytical tools from quantum theory but also potentially offers valuable insights into non-Hermitian physics.

In this paper, we study universal properties of the homogeneous disordered state in active matter models with uniaxial self-propulsion, such as variants of the active lattice gas model for MIPS~\cite{Thompson2011,Peruani2011,Whitelam2018,Kourbane-Houssene2018,Partridge2019,Mukherjee2024} and the active Ising model for flocking~\cite{Solon2013,Solon2015a}.
From hydrodynamic arguments, we propose that the density correlation and polarity correlation generically exhibit power-law decay with distinct exponents ($-2$ and $-4$, respectively) through the coupling of density and polarity inherent to active matter.
Performing simulations of lattice gas models with repulsive or aligning interaction, we numerically confirm that the predicted power-law correlations appear regardless of the interaction type.
For these models, employing the mapping to a quantum system and non-Hermitian perturbation theory, we analytically find that even the first order of the interaction strength induces the power-law correlations.

\section{Hydrodynamic argument}
\label{sec_hydro}

We consider a homogeneous disordered state of self-propelled particles in two dimensions, with the direction of self-propulsion restricted along the $x$ axis.
We assume that the particles flip their direction and move by diffusion or self-propulsion with interactions such as repulsion or alignment.
The macroscopic dynamics is expected to have the same form regardless of the interaction type and may be described~\footnote{Equation~\eqref{eq_hydro} is a simpler version of Eq.~(2) in Ref.~\cite{Adachi2022} [or Eq.~(S-21) in Ref.~\cite{Mukherjee2024}], which is derived for a repulsive active lattice gas model.} by a linear fluctuating hydrodynamic equation:
\begin{equation}
    \partial_t \rho_\pm = \mp v \partial_x \rho_\pm + \bar{D} \bm{\nabla}^2 \rho_\pm + \eta + \bar{\gamma} (\rho_\mp - \rho_\pm) \pm \xi,
    \label{eq_hydro}
\end{equation}
where $v, \bar{D}, \bar{\gamma} > 0$.
Here, $\rho_{\pm} (\bm{r}, t)$ is the coarse-grained density field for particles that are self-propelled in the $\pm x$ direction.
The first term on the right-hand side represents advection induced by self-propulsion with velocity $\pm v$.
The second and third terms describe stochastic diffusion in a standard way as used in the so-called Model B for dynamics of conserved density~\cite{Hohenberg1977,Chaikin1995}.
$\eta$ is a conserved Gaussian noise with $\braket{\eta (\bm{r}, t)}=0$, $\braket{\eta (\bm{r}, t) \eta (\bm{r}', t')} = -2 \Delta \bm{\nabla}^2 \delta (\bm{r} - \bm{r}') \delta (t - t')$, and $\Delta > 0$.
The last two terms denote flipping (i.e., rotation) of the direction of self-propulsion with rate $\bar{\gamma}$.
$\xi$ is a non-conserved Gaussian noise with $\braket{\xi (\bm{r}, t)} = 0$, $\braket{\xi (\bm{r}, t) \xi (\bm{r}', t')} = 2 \Delta' \delta (\bm{r} - \bm{r}') \delta (t - t')$, and $\Delta' > 0$.
The parameters $\bar{D}$ and $\bar{\gamma}$ quantify the strengths of diffusion and flipping, respectively, the values of which depend on the detail of the interaction.
The main conclusion will not change if we consider more complex noise correlations or further anisotropic terms (e.g., by replacing $\bar{D} \bm{\nabla}^2$ with $\bar{D}_x {\partial_x}^2 + \bar{D}_y {\partial_y}^2$).

Using the total density field $\rho (\bm{r}, t) := \rho_+ (\bm{r}, t) + \rho_- (\bm{r}, t)$ and the polarization density field $w (\bm{r}, t) := \rho_+ (\bm{r}, t) - \rho_- (\bm{r}, t)$~\cite{Marchetti2013}, we can rewrite Eq.~\eqref{eq_hydro} as
\begin{align}
    \partial_t \rho & = -v \partial_x w + \bar{D} \bm{\nabla}^2 \rho + 2 \eta
    \label{eq_hydro_rho}
    \\
    \partial_t w & = -v \partial_x \rho + \bar{D} \bm{\nabla}^2 w - 2 \bar{\gamma} w + 2 \xi.
\end{align}
Though we can directly solve these linear equations, we here use the adiabatic approximation~\cite{Speck2015} to reach the essential results in a simple way.
We notice that $\rho (\bm{r}, t)$ is a slow variable since the total particle number is conserved, while $w (\bm{r}, t)$ is a fast variable with decay rate $2 \bar{\gamma}$.
Thus, we use the adiabatic approximation as
\begin{equation}
    w \simeq -\frac{v}{2 \bar{\gamma}} \partial_x \rho + \frac{1}{\bar{\gamma}} \xi,
    \label{eq_w_adiabatic_approx}
\end{equation}
where we neglect higher-order gradient terms by considering long-wavelength fluctuations.
This equation indicates the coupling of density and polarity at the hydrodynamic level.
Substituting this into Eq.~\eqref{eq_hydro_rho}, we obtain
\begin{align}
    \partial_t \rho & \simeq \left[ \left( \bar{D} + \frac{v^2}{2 \bar{\gamma}} \right) {\partial_x}^2 + \bar{D} {\partial_y}^2 \right] \rho + \zeta
    \label{eq_hydro_approximate}
    \\
    \zeta & := 2 \eta - \frac{v}{\bar{\gamma}} \partial_x \xi
\end{align}
Due to the uniaxial self-propulsion with speed $v$, the obtained diffusion coefficient is anisotropic [i.e., $\bar{D} + v^2 / (2 \bar{\gamma})$ and $\bar{D}$ in the $x$ and $y$ directions, respectively], and the noise amplitude is also anisotropic, as seen from $\braket{\zeta (\bm{r}, t) \zeta (\bm{r}', t')} = -2 [(4 \Delta + v^2 \Delta' / \bar{\gamma}^2) {\partial_x}^2 + 4 \Delta {\partial_y}^2] \delta (\bm{r} - \bm{r}') \delta (t - t')$.

Solving Eq.~\eqref{eq_hydro_approximate}, we can obtain the steady-state density structure factor $\bar{S}_d (\bm{k}) := \Omega^{-1} \lim_{t \to \infty} \braket{|\delta \rho (\bm{k}, t)|^2}$, where $\delta \rho (\bm{k}, t) := \int d^2 \bm{r} e^{-i \bm{k} \cdot \bm{r}} [\rho (\bm{r}, t) - \bar{\rho}_0]$ with the average density $\bar{\rho}_0$, and $\Omega$ is the total area of the system.
For $\bm{k} \simeq \bm{0}$, we obtain
\begin{equation}
    \bar{S}_d (\bm{k}) \simeq \frac{(4 \Delta + v^2 \Delta' / \bar{\gamma}^2) {k_x}^2 + 4 \Delta {k_y}^2}{[\bar{D} + v^2 / (2 \bar{\gamma})] {k_x}^2 + \bar{D} {k_y}^2}.
\end{equation}
We notice that $\bar{S}_d (\bm{k})$ is generically discontinuous at $\bm{k} \to \bm{0}$, as characterized by
\begin{equation}
    \lim_{k_x \to 0} \bar{S}_d (k_x, 0) - \lim_{k_y \to 0} \bar{S}_d (0, k_y) = \frac{v^2 \Delta'}{{\bar{\gamma}}^2 \bar{D}} \frac{1 - 2 \bar{\gamma} \Delta / (\bar{D} \Delta')}{1 + v^2 / (2 \bar{\gamma} \bar{D})} \neq 0,
\end{equation}
except for the special case with $2 \bar{\gamma} \Delta = \bar{D} \Delta'$ where the fluctuation-dissipation relation happens to hold in Eq.~\eqref{eq_hydro_approximate}.

From the discontinuity of $\bar{S}_d (\bm{k})$, the steady-state density correlation function, $\bar{C}_d (\bm{r}) := \lim_{t \to \infty} \braket{\rho (\bm{0}, t) \rho (\bm{r}, t)} - {\bar{\rho}_0}^2 = \int d^2 \bm{k} e^{i \bm{k} \cdot \bm{r}} \bar{S}_d (\bm{k})$, shows power-law decay for $|\bm{r}| \to \infty$~\cite{Schmittmann1995}:
\begin{equation}
    \begin{split}
        \bar{C}_d (x, 0) & \sim x^{-2} \\
        \bar{C}_d (0, y) & \sim y^{-2}.
    \end{split}
    \label{eq_Cdbar_power_law}
\end{equation}
Note that the fluctuating hydrodynamic equation similar to Eq.~\eqref{eq_hydro_approximate} has been widely used to explain long-range density correlation of the form of Eq.~\eqref{eq_Cdbar_power_law} in anisotropic nonequilibrium systems such as driven lattice gas models~\cite{Schmittmann1995,Schmittmann1998}.
The authors have observed the same type of power-law density correlation in an active lattice gas model~\cite{Adachi2022} and active Brownian particles~\cite{Nakano2024} with uniaxial self-propulsion.

We further consider the steady-state polarity correlation function, $\bar{C}_p (\bm{r}) := \lim_{t \to \infty} \braket{w (\bm{0}, t) w (\bm{r}, t)}$.
Since $\bar{C}_p (\bm{r}) \sim {\partial_x}^2 \bar{C}_d (\bm{r})$ from Eq.~\eqref{eq_w_adiabatic_approx}, we obtain the power-law polarity correlation for $|\bm{r}| \to \infty$:
\begin{equation}
    \begin{split}
        \bar{C}_p (x, 0) & \sim x^{-4} \\
        \bar{C}_p (0, y) & \sim y^{-4},
    \end{split}
    \label{eq_Cpbar_power_law}
\end{equation}
where the exponent is decreased by $2$ compared to the density correlation.
Since we do not assume the interaction type in the above discussion, the power-law correlations of density [Eq.~\eqref{eq_Cdbar_power_law}] and polarity [Eq.~\eqref{eq_Cpbar_power_law}] are expected to appear in many-particle systems with uniaxial self-propulsion regardless of the detail of the interaction.

\section{Numetical simulations}
\label{sec_simulation}

\subsection{Lattice gas model with uniaxial self-propulsion}
\label{subsec_lattice_gas_model}

\begin{figure}[t]
    \centering
    \includegraphics[scale=1]{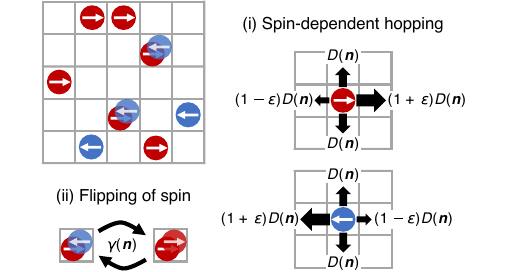}
    \caption{Lattice gas model with uniaxial self-propulsion.
    We consider $N$ particles with spin $+$ (red, right arrow) or $-$ (blue, left arrow) on a square lattice with size $L \times L$.
    Each particle stochastically (i) hops to the neighboring site at a rate favoring the direction of its spin or (ii) flips its spin.
    The hopping rate $D(\bm{n})$ and flipping rate $\gamma(\bm{n})$ depend on the particle configuration $\bm{n}$, representing the repulsive and aligning interactions between particles, respectively [see Eq.~\eqref{eq_lattice_model}].}
    \label{fig_model}
\end{figure}

To confirm the prediction [Eqs.~\eqref{eq_Cdbar_power_law} and \eqref{eq_Cpbar_power_law}] for uniaxially self-propelled particles with distinct interaction types, we consider a lattice gas model in two dimensions with lattice size $L \times L$, total particle number $N$, and periodic boundary conditions in both directions (Fig.~\ref{fig_model}).
Each particle has ``spin'' $s \in \{ +, - \}$ as an internal state, which corresponds to the direction of self-propulsion along the $x$ axis.
The particle configuration is specified by $\bm{n} := \{ n_{i, s} \}$, where $n_{i, s}$ is the number of particles with spin $s$ at site $i$.
A particle with spin $s$ follows two kinds of dynamics: (i) the particle hops one site right with rate $(1 + s \varepsilon) D (\bm{n})$, left with rate $(1 - s \varepsilon) D (\bm{n})$, and up or down with rate $D (\bm{n})$ for each, and (ii) the spin flips to $-s$ with rate $\gamma (\bm{n})$.
Here, $\varepsilon \in [0, 1]$ is a parameter for the strength of self-propulsion, where $\varepsilon = 0$ corresponds to no self-propulsion.
The interactions between particles are represented by the functional forms of $D (\bm{n})$ and $\gamma (\bm{n})$.

We consider repulsive and aligning interactions within Glauber-type dynamics:
\begin{equation}
    \begin{split}
        D (\bm{n}) & = D_0 \frac{2}{1 + e^{\Delta E_\mathrm{hop} (\bm{n})}} \\
        \gamma (\bm{n}) & = \gamma_0 \frac{2}{1 + e^{\Delta E_\mathrm{flip} (\bm{n})}},
        \label{eq_lattice_model}
    \end{split}
\end{equation}
where $D_0, \gamma_0 > 0$, and the functional forms of $\Delta E_\mathrm{hop} (\bm{n})$ and $\Delta E_\mathrm{flip} (\bm{n})$ specify the interactions.
For hopping from a site $i$ to a neighboring site $j$, we consider a repulsive interaction:
\begin{equation}
    \Delta E_\mathrm{hop} (\bm{n}) = U (n_j - n_i + 1),
    \label{eq_rep_interaction}
\end{equation}
where $U \geq 0$ is the strength of repulsion, and $n_i := \sum_s n_{i, s}$ is the local density.
For spin flipping from $s$ to $-s$, we consider an aligning interaction:
\begin{equation}
    \Delta E_\mathrm{flip} (\bm{n}) = J (s m_i - 1),
    \label{eq_al_interaction}
\end{equation}
where $J \geq 0$ is the strength of alignment, and $m_i := \sum_s s n_{i, s}$ is the local polarity.

\begin{figure}[t]
    \centering
    \includegraphics[scale=1]{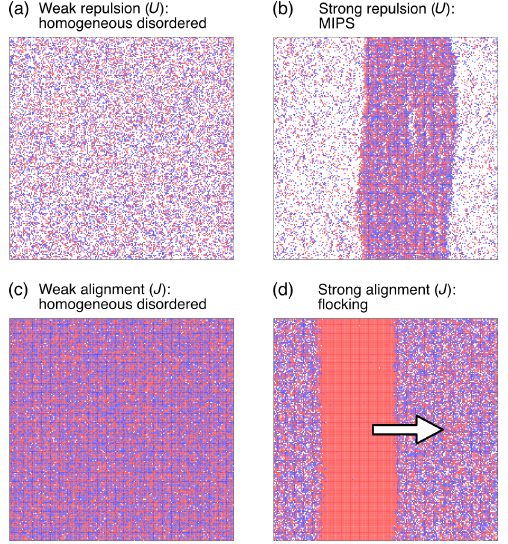}
    \caption{Interaction-dependent phase behavior.
    (a,~b) With repulsive interaction $U$, (a) the homogeneous disordered steady state appears for weak $U$, and (b) the system shows MIPS for strong $U$.
    (c,~d) With aligning interaction $J$, (c) the homogeneous disordered steady state appears for weak $J$, and (d) the system shows flocking for strong $J$.
    In each typical snapshot, the red, blue, and purple dots mean that the site is occupied by one or more particles with the local polarity $m_i$ positive, negative, and zero, respectively (see also Fig.~\ref{fig_model}).
    The arrow in (d) indicates the direction of the flocking motion.
    The used parameters are (a,~b) $D_0=1$, $\varepsilon=0.8$, $\gamma_0=0.01$, $L=200$, $N=2\times10^4$, and (a) $U=0.1$ or (b) $U=5$; (c,~d) $D_0=1$, $\varepsilon=0.8$, $\gamma_0=2$, $L=200$, $N=1.6\times10^5$, and (c) $J=0.1$ or (d) $J=1$.}
    \label{fig_phase}
\end{figure}

In Fig.~\ref{fig_phase}, we show the qualitative phase behavior of our model.
In the system with pure repulsion (i.e., $U > 0$ and $J = 0$), MIPS appears for large $U$ [Fig.~\ref{fig_phase}(b)], as observed in similar models~\cite{Kourbane-Houssene2018,Adachi2022,Mukherjee2024}.
The cluster configuration in MIPS reflects the anisotropy of self-propulsion.
In the system with pure alignment (i.e., $U = 0$ and $J > 0$), flocking appears for large $J$ [Fig.~\ref{fig_phase}(d)], as observed in the active Ising model~\cite{Solon2013,Solon2015a}.
The cluster configuration in flocking reflects the anisotropy of self-propulsion.
When $U$ and $J$ are small enough, noise overcomes the effects of interactions, and the steady state is homogeneous and disordered [Figs.~\ref{fig_phase}(a) and (c)].
In the following, we focus on the collective properties of these homogeneous disordered states.

For the lattice gas model, we define the density correlation function:
\begin{equation}
    C_d (\bm{r}_j) := \frac{1}{L^2} \sum_i \braket{n_i n_{i + j}} - {\rho_0}^2,
    \label{eq_def_Cd}
\end{equation}
where $\bm{r}_j$ is the coordinate of site $j$, $\braket{\cdots}$ is the steady-state ensemble average, and $\rho_0 := N / L^2$.
Similarly, we define the polarity correlation function:
\begin{equation}
    C_p (\bm{r}_j) := \frac{1}{L^2} \sum_i \braket{m_i m_{i + j}}.
    \label{eq_def_Cp}
\end{equation}

We also define the corresponding density and polarity structure factors:
\begin{equation}
    S_d (\bm{k}) := \sum_j e^{-i \bm{k} \cdot \bm{r}_j} C_d (\bm{r}_j)
\end{equation}
and
\begin{equation}
    S_p (\bm{k}) := \sum_j e^{-i \bm{k} \cdot \bm{r}_j} C_p (\bm{r}_j),
\end{equation}
where $\bm{k} := (2 \pi n_x / L, 2 \pi n_y / L) \in [-\pi, \pi) \times [-\pi, \pi)$ with $n_x, n_y \in \mathbb{Z}$.
Note that $S_d (\bm{0}) = 0$ from the particle number conservation (i.e., $\sum_i n_i = N$).

\begin{figure}[t]
    \centering
    \includegraphics[scale=1]{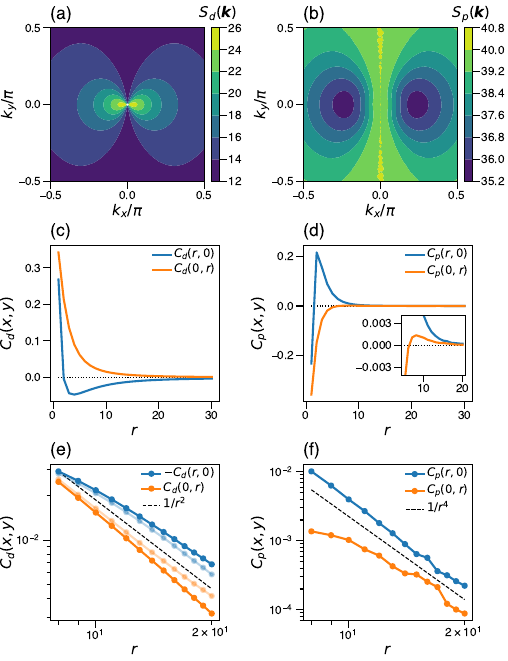}
    \caption{Power-law correlations in the system with repulsion and uniaxial self-propulsion.
    (a,~b) Heat maps of the structure factors for (a) density [$S_d (\bm{k})$] and (b) polarity [$S_p (\bm{k})$].
    (c,~d) Correlation functions along the $x$ axis (blue) and $y$ axis (orange) for (c) density [$C_d (\bm{r})$] and (d) polarity [$C_p (\bm{r})$].
    The inset of (d) is an expanded view, suggesting that $C_p(\bm{r})$ finally decays from the positive side in both $x$ and $y$ directions.
    (e,~f) Log-log plots of the (e) density and (f) polarity correlation functions, which correspond to (c) and (d), respectively.
    In (e), the two lines with light color suggest the correlation function after subtracting $-\rho_0/L^2$, which comes from a finite-size effect [see Eq.~\eqref{eq_Cd0}].
    The black dashed lines in (e) and (f) are algebraic functions, indicating that $C_d(\bm{r}) \sim r^{-2}$ and $C_p(\bm{r}) \sim r^{-4}$.
    The used parameters are $D_0 = 1$, $\varepsilon = 0.8$, $\gamma_0 = 0.5$, $U = 0.05$, $L = 200$, and $N = 1.6 \times 10^6$.}
    \label{fig_repulsion}
\end{figure}

\begin{figure}[t]
    \centering
    \includegraphics[scale=1]{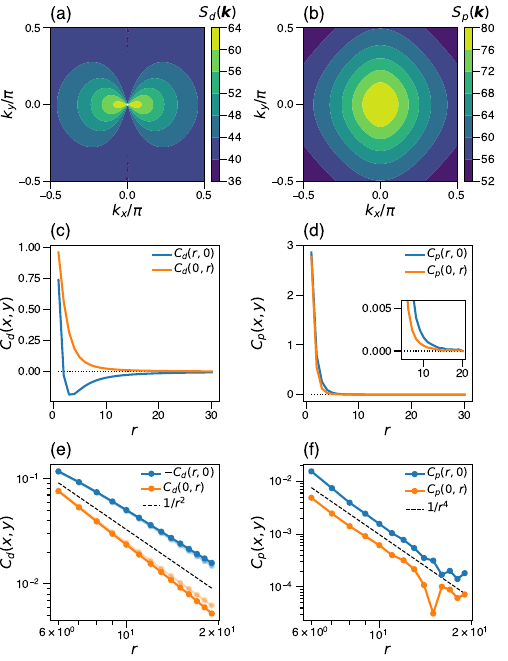}
    \caption{Power-law correlations in the system with alignment and uniaxial self-propulsion.
    We plot (a,~b) the structure factors and (c-f) the correlation functions in the same way as shown in Fig.~\ref{fig_repulsion}.
    In particular, (e) and (f) suggest that $C_d(\bm{r}) \sim r^{-2}$ and $C_p(\bm{r}) \sim r^{-4}$.
    The used parameters are $D_0 = 1$, $\varepsilon = 0.8$, $\gamma_0 = 2$, and $J = 0.025$, $L = 200$, and $N = 1.6 \times 10^6$.}
    \label{fig_alignment}
\end{figure}

\subsection{Power-law correlations for repulsive interaction}

We first consider the system with purely repulsive interaction (i.e., $U > 0$ and $J = 0$).
We set $D_0 = 1$, $\varepsilon = 0.8$, $\gamma_0 = 0.5$, $U = 0.05$, $L = 200$, and $N = 1.6 \times 10^6$, where the steady state is homogeneous and disordered.
We selected a high density, $\rho_0 = N / L^2 = 40$, to clearly see the asymptotic behavior of the correlation functions with a limited number of samples.
See Appendix~\ref{app_lattice_gas_model} for the simulation method and Appendix~\ref{app_sampling} for the numerical sampling of configurations in the steady state.

In Figs.~\ref{fig_repulsion}(a) and (b), we show the heat maps of the obtained density structure factor $S_d (\bm{k})$ and polarity structure factor $S_p (\bm{k})$, respectively.
$S_d (\bm{k})$ is discontinuous at $\bm{k} \to \bm{0}$, as expected from the hydrodynamic argument in Sec.~\ref{sec_hydro}.
Correspondingly, the density correlation function $C_d (\bm{r})$ exhibits the power-law decay with exponent $-2$ in the $x$ and $y$ directions [Figs.~\ref{fig_repulsion}(c) and (e)], which is consistent with Eq.~\eqref{eq_Cdbar_power_law}.
Lastly, the polarity correlation function $C_p (\bm{r})$ shows the power-law decay with exponent $-4$ [Figs.~\ref{fig_repulsion}(d) and (f)] as predicted in Eq.~\eqref{eq_Cpbar_power_law}.

\subsection{Power-law correlations for aligning interaction}

We next consider the system with purely aligning interaction (i.e., $U = 0$ and $J > 0$).
We set $D_0 = 1$, $\varepsilon = 0.8$, $\gamma_0 = 2$, and $J = 0.025$, $L = 200$, and $N = 1.6 \times 10^6$ (corresponding to $\rho_0 = 40$), where the steady state is homogeneous and disordered.
See Appendix~\ref{app_lattice_gas_model} for the simulation method and Appendix~\ref{app_sampling} for the numerical sampling of configurations in the steady state.

In Fig.~\ref{fig_alignment}, we arrange the figures in the same order as Fig.~\ref{fig_repulsion}, the case of repulsive interaction.
The observed density structure factor $S_d (\bm{k})$ [Fig.~\ref{fig_alignment}(a)] and density correlation function $C_d (\bm{r})$ [Figs.~\ref{fig_alignment}(c) and (e)] are qualitatively similar to those in the repulsive case [Figs.~\ref{fig_repulsion}(a), (c), and (e)].
Particularly, we observe the power-law decay with exponent $-2$ in the $x$ and $y$ directions, as predicted by the hydrodynamic argument [Eq.~\eqref{eq_Cdbar_power_law}].
The polarity structure factor $S_p (\bm{k})$ and correlation function $C_p (\bm{r})$ [Figs.~\ref{fig_alignment}(b) and (d)] exhibit distinct behaviors from those for the repulsive case [Figs.~\ref{fig_repulsion}(b) and (d)].
Specifically, polarity shows positive correlations in all directions [Fig.~\ref{fig_alignment}(d)] due to the aligning interaction, while the repulsive interaction leads to the negative polarity correlation at a short distance [Fig.~\ref{fig_repulsion}(d)].
Despite such difference in the short-distance property, the long-distance polarity correlation function $C_p (\bm{r})$ exhibits the power-law decay with exponent $-4$ in the $x$ and $y$ directions [Fig.~\ref{fig_alignment}(f)], which is also consistent with the hydrodynamic argument [Eq.~\eqref{eq_Cpbar_power_law}].

\section{Perturbation theory for interactions}

To show that very weak interaction, regardless of repulsion or alignment, is sufficient for the power-law correlations observed in the lattice gas model (Sec.~\ref{sec_simulation}), we analytically derive the correlation functions up to the first order in the interaction strength.
For the systematic derivation of the distribution of particle configurations, we map our model (Fig.~\ref{fig_model}) to a non-Hermitian quantum model by the Doi-Peliti method~\cite{Doi1976,Peliti1985,Tauber2014} and apply perturbation theory~\cite{Sternheim1972}.
With this approach, we can obtain analytical expressions of the structure factors without taking the hydrodynamic limit.
Readers who are not interested in the detailed derivation can skip to the main results in Sec.~\ref{subsec_corr_strfac}.

\subsection{Doi-Peliti method}

The master equation for our model is written as
\begin{equation}
    \frac{d}{d t} P (\bm{n}, t) = \sum_{\bm{n}'} W (\bm{n}, \bm{n}') P (\bm{n}', t),
    \label{eq_master_equation}
\end{equation}
where $P (\bm{n}, t)$ is the time-dependent distribution of the configuration $\bm{n}$, and $W (\bm{n}, \bm{n}')$ is the transition rate from $\bm{n}'$ to $\bm{n}$.
The steady-state distribution $P_\mathrm{st} (\bm{n})$ should satisfy
\begin{equation}
    \sum_{\bm{n}'} W (\bm{n}, \bm{n}') P_\mathrm{st} (\bm{n}') = 0.
\end{equation}
Regarding $W (\bm{n}, \bm{n}')$ as a matrix element of $W$, we can think of $P_\mathrm{st} (\bm{n})$ as a component of the eigenvector of $W$ corresponding to the eigenvalue of zero.

We introduce a boson Fock space spanned by Fock bases $\{ \ket{\bm{n}} \}_{\bm{n}}$.
Each Fock basis $\ket{\bm{n}}$ ($= \ket{\{ n_{i, s} \}}$) is defined using the vacuum state $\ket{0}$ as
\begin{equation}
    \ket{\bm{n}} := \prod_{i, s} \left( \hat{a}_{i, s}^\dag \right)^{n_{i, s}} \ket{0},
\end{equation}
and the corresponding left Fock basis is defined as
\begin{equation}
    {}_L \! \bra{\bm{n}} := \bra{0} \prod_{i, s} \frac{1}{n_{i, s}!} \left( \hat{a}_{i, s} \right)^{n_{i, s}}.
\end{equation}
Here, $\hat{a}_{i, s}^{(\dag)}$ is the annihilation (creation) operator of a boson with spin $s$ at site $i$, and we assume the commutation relation: $[\hat{a}_{i, s}, \hat{a}_{j, s'}^\dag] = \delta_{i, j} \delta_{s, s'}$ and $[\hat{a}_{i, s}, \hat{a}_{j, s'}] = [\hat{a}_{i, s}^\dag, \hat{a}_{j, s'}^\dag] = 0$.
The number operator of a boson with spin $s$ at site $i$ is given as $\hat{n}_{i, s} := \hat{a}_{i, s}^\dag \hat{a}_{i, s}$.
Note that $\ket{\bm{n}}$ and ${}_L {\bra{\bm{n}}}$ satisfy biorthogonality: ${}_L \! \braket{\bm{n} | \bm{n}'} = \delta_{\bm{n}, \bm{n}'}$.

We define the Fourier transformation of the operator $\hat{a}_{j, s}$ for wavevector $\bm{k} = (2 \pi n_x / L, 2 \pi n_y / L)$ with $n_x, n_y \in \mathbb{Z}$ and $k_x, k_y \in [-\pi, \pi)$:
\begin{equation}
    \hat{a}_{\bm{k}, s} := \frac{1}{L} \sum_j e^{-i \bm{k} \cdot \bm{r}_j} \hat{a}_{j, s},
    \label{eq_aks_def}
\end{equation}
where $\bm{r}_j$ is the coordinate of site $j$.
The operators satisfy the boson commutation relation: $[\hat{a}_{\bm{k}, s}, \hat{a}_{\bm{k}', s'}^\dag] = \delta_{\bm{k}, \bm{k}'} \delta_{s, s'}$ and $[\hat{a}_{\bm{k}, s}, \hat{a}_{\bm{k}', s'}] = [\hat{a}_{\bm{k}, s}^\dag, \hat{a}_{\bm{k}', s'}^\dag] = 0$.
The inverse transformation is given as
\begin{equation}
    \hat{a}_{j, s} = \frac{1}{L} \sum_{\bm{k}} e^{i \bm{k} \cdot \bm{r}_j} \hat{a}_{\bm{k}, s}.
    \label{eq_ajs_inv_transformation}
\end{equation}

We define the time-dependent state vector:
\begin{equation}
    \ket{\psi (t)} := \sum_{\bm{n}} P(\bm{n}, t) \ket{\bm{n}}.
\end{equation}
Then, the master equation [Eq.~\eqref{eq_master_equation}] is equivalent to the following imaginary-time Schr{\" o}dinger equation~\cite{Tauber2014}:
\begin{equation}
    \frac{d}{d t} \ket{\psi (t)} = -\hat{H} \ket{\psi (t)},
\end{equation}
where the matrix element of the pseudo-Hamiltonian $\hat{H}$ is defined by
\begin{equation}
    {}_L \! \braket{\bm{n} | \hat{H} | \bm{n}'} := -W (\bm{n}, \bm{n}').
    \label{eq_def_H}
\end{equation}
Thus, the steady-state vector,
\begin{equation}
    \ket{\psi_\mathrm{st}} := \sum_{\bm{n}} P_\mathrm{st} (\bm{n}) \ket{\bm{n}},
\end{equation}
is the eigenvector of $\hat{H}$ corresponding to the eigenvalue of zero.
Since $P_\mathrm{st} (\bm{n}) = {}_L \! \braket{\bm{n} | \psi_\mathrm{st}}$, the calculation of the steady-state distribution $P_\mathrm{st} (\bm{n})$ is equivalent to the eigenvalue problem of $\hat{H}$ (i.e., calculation of $\ket{\psi_\mathrm{st}}$).

The steady-state average of a configuration-dependent physical quantity $A(\bm{n})$ can be expressed~\cite{Tauber2014} as
\begin{equation}
    \braket{A (\bm{n})} = \sum_{\bm{n}} A(\bm{n}) P_\mathrm{st} (\bm{n}) = \braket{\mathcal{P} | A({\hat{\bm{n}}}) | \psi_\mathrm{st}},
    \label{eq_average_formula}
\end{equation}
where $\hat{\bm{n}} := \{ \hat{n}_{i, s} \}$ and
\begin{equation}
    \bra{\mathcal{P}} := \bra{0} e^{\sum_{i, s} \hat{a}_{i, s}}.
    \label{eq_projection_def}
\end{equation}
Note that $\sum_{\bm{n}} P_\mathrm{st} (\bm{n}) = 1$ is equivalent to $\braket{\mathcal{P} | \psi_\mathrm{st}} = 1$, as obtained from Eq.~\eqref{eq_average_formula} by setting $A (\bm{n}) = 1$.

Using this formulation, we can write the density correlation function [Eq.~\eqref{eq_def_Cd}] as
\begin{align}
    C_d (\bm{r}_j) & = \frac{1}{L^2} \sum_i \braket{n_i n_{i + j}} - {\rho_0}^2 \nonumber \\
    & = \frac{1}{L^2} \sum_i \braket{\mathcal{P} | \hat{n}_i \hat{n}_{i + j} | \psi_\mathrm{st}} - {\rho_0}^2 \nonumber \\
    & = \frac{1}{L^2} \sum_i \sum_{s, s'} \braket{\mathcal{P} | \hat{a}_{i, s} \hat{a}_{i + j, s'} | \psi_\mathrm{st}} + \rho_0 \delta_{\bm{r}_j, \bm{0}} - {\rho_0}^2,
\end{align}
where $\hat{n}_i := \sum_s \hat{n}_{i, s}$, $\rho_0 = N / L^2$, and we used $\bra{\mathcal{P}} \hat{a}_{i, s}^\dag = \bra{\mathcal{P}}$.
Similarly, the polarity correlation function [Eq.~\eqref{eq_def_Cp}] is obtained as
\begin{align}
    C_p (\bm{r}_j) & = \frac{1}{L^2} \sum_i \braket{m_i m_{i + j}} \nonumber \\
    & = \frac{1}{L^2} \sum_i \braket{\mathcal{P} | \hat{m}_i \hat{m}_{i + j} | \psi_\mathrm{st}} \nonumber \\
    & = \frac{1}{L^2} \sum_i \sum_{s, s'} s s' \braket{\mathcal{P} | \hat{a}_{i, s} \hat{a}_{i + j, s'} | \psi_\mathrm{st}} + \rho_0 \delta_{\bm{r}_j, \bm{0}},
\end{align}
where $\hat{m}_i := \sum_s s \hat{n}_{i, s}$.

We also obtain the corresponding density and polarity structure factors as
\begin{align}
    S_d (\bm{k}) & = \sum_j e^{-i \bm{k} \cdot \bm{r}_j} C_d (\bm{r}_j) \nonumber \\
    & = \sum_{s, s'} \braket{\mathcal{P} | \hat{a}_{\bm{k}, s} \hat{a}_{-{\bm{k}, s'}} | \psi_\mathrm{st}} + \rho_0 - {\rho_0}^2 L^2 \delta_{\bm{k}, \bm{0}}
    \label{eq_Sd_def}
\end{align}
and
\begin{align}
    S_p (\bm{k}) & = \sum_j e^{-i \bm{k} \cdot \bm{r}_j} C_p (\bm{r}_j) \nonumber \\
    & = \sum_{s, s'} s s' \braket{\mathcal{P} | \hat{a}_{\bm{k}, s} \hat{a}_{-{\bm{k}, s'}} | \psi_\mathrm{st}} + \rho_0,
\end{align}
respectively.

\subsection{Pseudo-Hamiltonian}

We divide $\hat{H}$ [Eq.~\eqref{eq_def_H}] into two parts:
\begin{equation}
    \hat{H} = \hat{H}_0 + \hat{H}_\mathrm{int},
\end{equation}
where $\hat{H}_0$ is the unperturbed (i.e., non-interacting) part, and $\hat{H}_\mathrm{int}$ is the perturbation (i.e., interaction) part.

$\hat{H}_0$ is obtained as
\begin{align}
    \hat{H}_0 & = - D_0 \sum_{\braket{i, j}, s} \hat{a}_{j, s}^\dag \hat{a}_{i, s} - \varepsilon D_0 \sum_{i, s} s (\hat{a}_{i + \hat{x}, s}^\dag \hat{a}_{i, s} - \hat{a}_{i - \hat{x}, s}^\dag \hat{a}_{i, s}) \nonumber \\
    & \ \ \ \ - \gamma_0 \sum_{i, s} \hat{a}_{i, -s}^\dag \hat{a}_{i, s} + (4 D_0 + \gamma_0) N,
    \label{eq_def_H0}
\end{align}
where $\sum_{\braket{i, j}}$ denotes the summation over all pairs of neighboring sites $(i, j)$, with the pairs $(i, j)$ and $(j, i)$ being treated as distinct, and $\hat{x}$ is the unit translation along the $x$ axis.
The first and third terms are Hermitian and represent the symmetric hopping and spin flipping, respectively; the second term is anti-Hermitian and represents self-propulsion; the last term is the diagonal part of $\hat{H}_0$, which is a constant due to the conservation of the total particle number $N$.

$\hat{H}_\mathrm{int}$ is obtained as
\begin{align}
    \hat{H}_\mathrm{int} & = D_0 \sum_{\braket{i, j}, s} (\hat{a}_{j, s}^\dag \hat{a}_{i, s} - \hat{n}_{i, s}) \tanh \left[ \frac{U (\hat{n}_j - \hat{n}_i + 1)}{2} \right] \nonumber \\
    & \ \ \ \ + \varepsilon D_0 \sum_{i, s} s (\hat{a}_{i + \hat{x}, s}^\dag \hat{a}_{i, s} - \hat{n}_{i, s}) \tanh \left[ \frac{U (\hat{n}_{i + \hat{x}} - \hat{n}_i + 1)}{2} \right] \nonumber \\
    & \ \ \ \ - \varepsilon D_0 \sum_{i, s} s (\hat{a}_{i - \hat{x}, s}^\dag \hat{a}_{i, s} - \hat{n}_{i, s}) \tanh \left[ \frac{U (\hat{n}_{i - \hat{x}} - \hat{n}_i + 1)}{2} \right] \nonumber \\
    & \ \ \ \ + \gamma_0 \sum_{i, s} (\hat{a}_{i, -s}^\dag \hat{a}_{i, s} - \hat{n}_{i, s}) \tanh \left[ \frac{J (s \hat{m}_i - 1)}{2} \right].
    \label{eq_Hint}
\end{align}
Here, the first three lines and the last line represent the effects of repulsion and alignment [see Eqs.~\eqref{eq_lattice_model}-\eqref{eq_al_interaction}], respectively.
Note that $\hat{H}_\mathrm{int}$ is generically non-Hermitian.

\subsection{Non-Hermitian perturbation theory}

Considering small $U$ and $J$, we expand $\hat{H}_\mathrm{int}$ [Eq.~\eqref{eq_Hint}] up to the first order in terms of $U$ and $J$:
\begin{align}
    \hat{H}_\mathrm{int} \simeq \hat{H}_\mathrm{int}' & := \frac{U D_0} {2} \sum_{\braket{i, j}, s} (\hat{a}_{j, s}^\dag \hat{a}_{i, s} - \hat{n}_{i, s}) (\hat{n}_j - \hat{n}_i + 1) \nonumber \\
    & \ \ \ \ + \frac{\varepsilon U D_0} {2} \sum_{i, s} s (\hat{a}_{i + \hat{x}, s}^\dag \hat{a}_{i, s} - \hat{n}_{i, s}) (\hat{n}_{i + \hat{x}} - \hat{n}_i + 1) \nonumber \\
    & \ \ \ \ - \frac{\varepsilon U D_0} {2} \sum_{i, s} s (\hat{a}_{i - \hat{x}, s}^\dag \hat{a}_{i, s} - \hat{n}_{i, s}) (\hat{n}_{i - \hat{x}} - \hat{n}_i + 1) \nonumber \\
    & \ \ \ \ + \frac{J \gamma_0}{2} \sum_{i, s} (a_{i, -s}^\dag a_{i, s} - \hat{n}_{i, s}) (s \hat{m}_i - 1).
    \label{eq_Hint_approximate}
\end{align}
Then, we expand the steady-state vector $\ket{\psi_\mathrm{st}}$ up to the first order in terms of $\hat{H}_\mathrm{int}'$:
\begin{equation}
    \ket{\psi_\mathrm{st}} \simeq \ket{\psi_\mathrm{st}^{(0)}} + \ket{\psi_\mathrm{st}^{(1)}},
\end{equation}
where $\ket{\psi_\mathrm{st}^{(0)}}$ is the steady-state vector of $\hat{H}_0$, and $\ket{\psi_\mathrm{st}^{(1)}}$ is the first-order correction.
Applying perturbation theory for non-Hermitian systems~\cite{Sternheim1972}, we obtain
\begin{equation}
    \ket{\psi_\mathrm{st}^{(1)}} = -\sum_{n \neq 0} \frac{{}_L \! \braket{\psi^{(0)} (n) | \hat{H} _\mathrm{int}' | \psi_\mathrm{st}^{(0)}}}{E^{(0)} (n)} \ket{\psi^{(0)} (n)}.
    \label{eq_ssvector_first_order_formula}
\end{equation}
Here, $\ket{\psi^{(0)} (n)}$, ${}_L \! \bra{\psi^{(0)} (n)}$, and $E^{(0)} (n)$ are the eigenvector, left eigenvector, and eigenvalue of $\hat{H}_0$ specified by $n$, respectively, and $n = 0$ corresponds to the steady state [e.g., $\ket{\psi^{(0)} (0)} = \ket{\psi_\mathrm{st}^{(0)}}$].
We assume biorthogonality: ${}_L \! \braket{\psi^{(0)} (n) | \psi^{(0)} (m)} = \delta_{n, m}$.

To calculate $\ket{\psi_\mathrm{st}^{(1)}}$ using Eq.~\eqref{eq_ssvector_first_order_formula}, we need to obtain $\ket{\psi_\mathrm{st}^\mathrm{(0)}}$ and $\{ \ket{\psi^{(0)} (n)}, {}_L \! \bra{\psi^{(0)} (n)}, E^{(0)} (n) \}$ for all $n$ by solving the eigenvalue problem for $\hat{H}_0$.
In the following, we employ linear transformations of the annihilation and creation operators that preserve the boson commutation relation.

From Eqs.~\eqref{eq_ajs_inv_transformation} and \eqref{eq_def_H0}, we can express $\hat{H}_0$ as
\begin{equation}
    \hat{H}_0 = \sum_{\bm{k}} 
    \begin{pmatrix}
        \hat{a}_{\bm{k}, +}^\dag & \hat{a}_{\bm{k}, -}^\dag
    \end{pmatrix}
    h_{\bm{k}}
    \begin{pmatrix}
        \hat{a}_{\bm{k}, +} \\
        \hat{a}_{\bm{k}, -}
    \end{pmatrix}
    ,
    \label{eq_block_diagonalized_H0}
\end{equation}
where
\begin{widetext}
    \begin{equation}
        h_{\bm{k}} =
        \begin{pmatrix}
            2 D_0 (2 - \cos k_x - \cos k_y + i \varepsilon \sin k_x) + \gamma_0 & -\gamma_0 \\
            -\gamma_0 & 2 D_0 (2 - \cos k_x - \cos k_y - i \varepsilon \sin k_x) + \gamma_0
        \end{pmatrix}
        .
    \end{equation}
\end{widetext}
Solving the eigenvalue problem of $2 \times 2$ matrix $h_{\bm{k}}$, we obtain the eigenvector $\bm{u}_{\bm{k}, \alpha}$, left eigenvector $\bm{v}_{\bm{k}, \alpha}^T$, and eigenvalue $\epsilon_{\bm{k}, \alpha}$ ($\alpha \in \{ 0, 1 \}$, $h_{\bm{k}} \bm{u}_{\bm{k}, \alpha} = \epsilon_{\bm{k}, \alpha} \bm{u}_{\bm{k}, \alpha}$, and $\bm{v}_{\bm{k}, \alpha}^T h_{\bm{k}} = \bm{v}_{\bm{k}, \alpha}^T \epsilon_{\bm{k}, \alpha}$):
\begin{align}
    \bm{u}_{\bm{k}, 0} & = \frac{1}{\sqrt{2}}
    \begin{pmatrix}
        1 \\
        \sqrt{1 - {f_{\bm{k}}}^2} + i f_{\bm{k}}
    \end{pmatrix}
    \label{eq_uk0}
    \\
    \bm{u}_{\bm{k}, 1} & = \frac{1}{\sqrt{2}}
    \begin{pmatrix}
        1 \\
        -\sqrt{1 - {f_{\bm{k}}}^2} + i f_{\bm{k}}
    \end{pmatrix}
    ,
\end{align}
\begin{align}
    \bm{v}_{\bm{k}, 0} & = \frac{1}{\sqrt{2}} \frac{1}{\sqrt{1 - {f_{\bm{k}}}^2}}
    \begin{pmatrix}
        \sqrt{1 - {f_{\bm{k}}}^2} - i f_{\bm{k}} \\ 1
    \end{pmatrix}
    \label{eq_vk0}
    \\
    \bm{v}_{\bm{k}, 1} & = \frac{1}{\sqrt{2}} \frac{1}{\sqrt{1 - {f_{\bm{k}}}^2}}
    \begin{pmatrix}
        \sqrt{1 - {f_{\bm{k}}}^2} + i f_{\bm{k}} \\ -1
    \end{pmatrix}
    ,
\end{align}
and
\begin{align}
    \epsilon_{\bm{k}, 0} & = 2 D_0 (2 - \cos k_x - \cos k_y) + \gamma_0 (1 - \sqrt{1 - {f_{\bm{k}}}^2})
    \\
    \epsilon_{\bm{k}, 1} & = 2 D_0 (2 - \cos k_x - \cos k_y) + \gamma_0 (1 + \sqrt{1 - {f_{\bm{k}}}^2}),
    \label{eq_epsilonk1}
\end{align}
where $f_{\bm{k}} := (2 \varepsilon D_0 / \gamma_0) \sin k_x$ and $\sqrt{1 - {f_{\bm{k}}}^2} = i \sqrt{{f_{\bm{k}}}^2 - 1}$ for $|f_{\bm{k}}| > 1$.
Note that $\bm{u}_{\bm{k}, \alpha}$ and $\bm{v}_{\bm{k}, \alpha}^T$ satisfy biorthogonality: $\bm{v}_{\bm{k}, \alpha}^T \bm{u}_{\bm{k}', \alpha'} = \delta_{\bm{k}, \bm{k}'} \delta_{\alpha, \alpha'}$.

We define a linear transformation of $\hat{a}_{\bm{k}, s}$ and $\hat{a}_{\bm{k}, s}^\dag$ as
\begin{align}
    \hat{b}_{\bm{k}, \alpha} & := \bm{v}_{\bm{k}, \alpha}^T
    \begin{pmatrix}
        \hat{a}_{\bm{k}, +} \\
        \hat{a}_{\bm{k}, -}
    \end{pmatrix}
    \label{eq_bkalpha_def}
    \\
    \hat{\bar{b}}_{\bm{k}, \alpha} & :=
    \begin{pmatrix}
        \hat{a}_{\bm{k}, +}^\dag & \hat{a}_{\bm{k}, -}^\dag
    \end{pmatrix}
    \bm{u}_{\bm{k}, \alpha}.
    \label{eq_bkalphabar_def}
\end{align}
From the biorthogonality of $\bm{u}_{\bm{k}, \alpha}$ and $\bm{v}_{\bm{k}, \alpha}^T$, the inverse transformation is given as
\begin{align}
    \begin{pmatrix}
        \hat{a}_{\bm{k}, +} \\
        \hat{a}_{\bm{k}, -}
    \end{pmatrix}
    = \sum_\alpha \bm{u}_{\bm{k}, \alpha} \hat{b}_{\bm{k}, \alpha}
    \label{eq_aks_inv_transformation}
    \\
    \begin{pmatrix}
        \hat{a}_{\bm{k}, +}^\dag & \hat{a}_{\bm{k}, -}^\dag
    \end{pmatrix}
    = \sum_\alpha \hat{\bar{b}}_{\bm{k}, \alpha} \bm{v}_{\bm{k}, \alpha}^T.
    \label{eq_aksdag_inv_transformation}
\end{align}
Note that $\hat{\bar{b}}_{\bm{k}, \alpha} \neq \hat{b}_{\bm{k}, \alpha}^\dag$ in general, except the case with no self-propulsion (i.e., $\varepsilon = 0$), where $f_{\bm{k}} = 0$ and $\bm{u}_{\bm{k}, \alpha} = \bm{v}_{\bm{k}, \alpha}$.
However, since the commutation relation holds (i.e., $[\hat{b}_{\bm{k}, \alpha}, \hat{\bar{b}}_{\bm{k}', \alpha'}] = \delta_{\bm{k}, \bm{k}'} \delta_{\alpha, \alpha'}$ and $[\hat{b}_{\bm{k}, \alpha}, \hat{b}_{\bm{k}', \alpha'}] = [\hat{\bar{b}}_{\bm{k}, \alpha}, \hat{\bar{b}}_{\bm{k}', \alpha'}] = 0$), $\hat{b}_{\bm{k}, \alpha}$, $\hat{\bar{b}}_{\bm{k}, \alpha}$, and $\hat{\bar{b}}_{\bm{k}, \alpha} \hat{b}_{\bm{k}, \alpha}$ are the annihilation, creation, and number operators for a quasiparticle specified by $(\bm{k}, \alpha)$.

Using the spectral decomposition $h_{\bm{k}} = \sum_\alpha \epsilon_{\bm{k}, \alpha} \bm{u}_{\bm{k}, \alpha} \bm{v}_{\bm{k}, \alpha}^T$ in Eq.~\eqref{eq_block_diagonalized_H0}, we obtain
\begin{equation}
    \hat{H}_0 = \sum_{\bm{k}, \alpha} \epsilon_{\bm{k}, \alpha} \hat{\bar{b}}_{\bm{k}, \alpha} \hat{b}_{\bm{k}, \alpha}.
    \label{eq_diagonalized_H0}
\end{equation}
Thus, any eigenvector of $\hat{H}_0$ is specified by the set of occupancies for all quasiparticle states, which is denoted by $\{ n_{\bm{k}, \alpha} \}$.
The eigenvector $\ket{\psi^{(0)} (\{ n_{\bm{k}, \alpha} \})}$ is given as
\begin{equation}
    \ket{\psi^{(0)} (\{ n_{\bm{k}, \alpha} \})} = \prod_{\bm{k}, \alpha} \left( \hat{\bar{b}}_{\bm{k}, \alpha} \right)^{n_{\bm{k}, \alpha}} \ket{0}.
    \label{eq_psi0}
\end{equation}
The corresponding left eigenvector ${}_L \! \bra{\psi^{(0)} (\{ n_{\bm{k}, \alpha} \})}$ and eigenvalue $E^{(0)} (\{ n_{\bm{k}, \alpha} \})$ are
\begin{equation}
    {}_L \! \bra{\psi^{(0)} (\{ n_{\bm{k}, \alpha} \})} = \bra{0} \prod_{\bm{k}, \alpha} \frac{1}{n_{\bm{k}, \alpha} !} \left( \hat{b}_{\bm{k}, \alpha} \right)^{n_{\bm{k}, \alpha}}
\end{equation}
and
\begin{equation}
    E^{(0)} (\{ n_{\bm{k}, \alpha} \}) = \sum_{\bm{k}, \alpha} \epsilon_{\bm{k}, \alpha} n_{\bm{k}, \alpha},
    \label{eq_E0}
\end{equation}
respectively.
Note that $\sum_{\bm{k}, \alpha} n_{\bm{k}, \alpha} = N$ since $\sum_{\bm{k}, \alpha} \hat{\bar{b}}_{\bm{k}, \alpha} \hat{b}_{\bm{k}, \alpha} = \sum_{i, s} \hat{a}_{i, s}^\dag \hat{a}_{i, s} = N$ is the fixed total particle number.

From Eqs.~\eqref{eq_psi0} and \eqref{eq_E0}, we find that the steady-state vector of $\hat{H}_0$, which corresponds to the eigenvalue of zero [i.e., $E^{(0)} (\{ n_{\bm{k}, \alpha} \}) = 0$], is obtained by taking $n_{\bm{0}, 0} = N$ and $n_{\bm{k}, \alpha} = 0$ for all $(\bm{k}, \alpha) \neq (\bm{0}, 0)$:
\begin{equation}
    \ket{\psi_\mathrm{st}^{(0)}} = \frac{1}{(\sqrt{2} L)^N} \left( \hat{\bar{b}}_{\bm{0}, 0} \right)^{N} \ket{0},
    \label{eq_psist0_explicit}
\end{equation}
where the normalization factor $(\sqrt{2} L)^{-N}$ ensures that the total probability for the configuration is $1$ (i.e., $\braket{\mathcal{P} | \psi_\mathrm{st}^{(0)}} = \sum_{\bm{n}} {}_L \! \braket{\bm{n} | \psi_\mathrm{st}^{(0)}} = 1$) [see the notes after Eq.~\eqref{eq_projection_def}].
This steady-state vector represents the random distribution of $N$ particles with no spatial or spin correlations since $\hat{\bar{b}}_{\bm{0}, 0} \propto \sum_{i, s} \hat{a}_{i, s}^\dag$, which follows from Eqs.~\eqref{eq_aks_def}, \eqref{eq_uk0}, and \eqref{eq_bkalphabar_def}.
All the other eigenvectors are specified by the set of occupancies $\{ n_{\bm{k}, \alpha} \}$ with $n_{\bm{0}, 0} < N$.
Thus, $\ket{\psi_\mathrm{st}^{(1)}}$ in Eq.~\eqref{eq_ssvector_first_order_formula} is expressed as
\begin{equation}
    \ket{\psi_\mathrm{st}^{(1)}} = -\sideset{}{'}{\sum}_{\{ n_{\bm{k}, \alpha} \}} \frac{{}_L \! \braket{\psi^{(0)} (\{ n_{\bm{k}, \alpha} \}) | \hat{H} _\mathrm{int}' | \psi_\mathrm{st}^{(0)}}}{E^{(0)} (\{ n_{\bm{k}, \alpha} \})} \ket{\psi^{(0)} (\{ n_{\bm{k}, \alpha} \})},
    \label{eq_first_order_ss_vector}
\end{equation}
where $\sum_{\{ n_{\bm{k}, \alpha} \}}'$ denotes the summation over all sets of occupancies $\{ n_{\bm{k}, \alpha} \}$ satisfying $\sum_{\bm{k}, \alpha} n_{\bm{k}, \alpha} = N$ and $n_{\bm{0}, 0} < N$.

To reduce Eq.~\eqref{eq_first_order_ss_vector} to a more explicit formula, we rewrite $\hat{H}_\mathrm{int}'$ [Eq.~\eqref{eq_Hint_approximate}] using $\hat{b}_{\bm{k}, \alpha}$ and $\hat{\bar{b}}_{\bm{k}, \alpha}$.
From Eqs.~\eqref{eq_ajs_inv_transformation}, \eqref{eq_Hint_approximate}, \eqref{eq_aks_inv_transformation}, and \eqref{eq_aksdag_inv_transformation}, we obtain
\begin{align}
    \hat{H}_\mathrm{int}' & = \frac{1}{L^2} \! \! \! \! \sum_{\substack{\bm{k}, \bm{k}', \bm{q} \\ s_1, s_2, s_3, s_4}} \! \! \! \! \! V_{s_1, s_2, s_3, s_4} (\bm{k}, \bm{k}', \bm{q}) \hat{a}_{\bm{q} + \bm{k}, s_1}^\dag \hat{a}_{\bm{q} - \bm{k}, s_2}^\dag \hat{a}_{\bm{q} - \bm{k}', s_4} \hat{a}_{\bm{q} + \bm{k}', s_3} \nonumber \\
    & = \frac{1}{L^2} \! \! \! \! \! \! \sum_{\substack{\bm{k}, \bm{k}', \bm{q} \\ \alpha_1, \alpha_2, \alpha_3, \alpha_4}} \! \! \! \! \! \! \tilde{V}_{\alpha_1, \alpha_2, \alpha_3, \alpha_4} (\bm{k}, \bm{k}', \bm{q}) \hat{\bar{b}}_{\bm{q} + \bm{k}, \alpha_1}  \hat{\bar{b}}_{\bm{q} - \bm{k}, \alpha_2} \hat{b}_{\bm{q} - \bm{k}', \alpha_4} \hat{b}_{\bm{q} + \bm{k}', \alpha_3},
    \label{eq_Hint_approximate_Fourier}
\end{align}
where
\begin{widetext}
    \begin{align}
        V_{s_1, s_2, s_3, s_4} (\bm{k}, \bm{k}', \bm{q}) &= U D_0 \delta_{s_1, s_3} \delta_{s_2, s_4} [2 - \cos (q_x + k_x) + \cos (q_x + k_x') - \cos (k_x - k_x') - \cos (q_y + k_y) + \cos (q_y + k_y') - \cos (k_y - k_y')] \nonumber \\
        & \ \ \ \ + i s_3 \varepsilon U D_0 \delta_{s_1, s_3} \delta_{s_2, s_4} [\sin (q_x + k_x) - \sin (q_x + k_x') - \sin (k_x - k_x')] + \frac{J \gamma_0}{2} s_3 s_4 (\delta_{s_1, -s_3} - \delta_{s_1, s_3}) \delta_{s_2, s_4} \\
        \tilde{V}_{\alpha_1, \alpha_2, \alpha_3, \alpha_4} (\bm{k}, \bm{k}', \bm{q}) &= \sum_{s_1, s_2, s_3, s_4} V_{s_1, s_2, s_3, s_4} (\bm{k}, \bm{k}', \bm{q}) v_{\bm{q} + \bm{k}, \alpha_1, s_1} v_{\bm{q} - \bm{k}, \alpha_2, s_2} u_{\bm{q} - \bm{k}', \alpha_4, s_4} u_{\bm{q} + \bm{k}', \alpha_3, s_3}.
        \label{eq_Vtilde}
    \end{align}
\end{widetext}
Note that $u_{\bm{k}, \alpha, s}$ and $v_{\bm{k}, \alpha, s}$ denote the components of $\bm{u}_{\bm{k}, \alpha}$ and $\bm{v}_{\bm{k}, \alpha}$, respectively.

According to the forms of $\ket{\psi_\mathrm{st}^{(0)}}$ [Eq.~\eqref{eq_psist0_explicit}] and $\hat{H}_\mathrm{int}'$ [Eq.~\eqref{eq_Hint_approximate_Fourier}], the first-order perturbation [Eq.~\eqref{eq_first_order_ss_vector}] causes two-quasiparticle excitations in the language of quantum theory.
From Eqs.~\eqref{eq_psi0}-\eqref{eq_Hint_approximate_Fourier}, we obtain
\begin{equation}
    \ket{\psi_\mathrm{st}^{(1)}} = -\frac{N (N - 1)}{(\sqrt{2} L)^N L^2} \sideset{}{'}{\sum}_{\bm{k}, \alpha_1, \alpha_2} \frac{\tilde{V}_{\alpha_1, \alpha_2, 0, 0} (\bm{k}, \bm{0}, \bm{0})}{\epsilon_{\bm{k}, \alpha_1} + \epsilon_{-\bm{k}, \alpha_2}} \ket{\bm{k}, \alpha_1; -\bm{k}, \alpha_2}.
    \label{eq_psist1_explicit}
\end{equation}
Here, $\sum_{\bm{k}, \alpha_1, \alpha_2}'$ denotes the summation over all sets of $(\bm{k}, \alpha_1, \alpha_2) \notin \{ (\bm{0}, 0, 0), (\bm{0}, 0, 1), (\bm{0}, 1, 0) \}$, and
\begin{equation}
    \ket{\bm{k}, \alpha_1; -\bm{k}, \alpha_2} := \hat{\bar{b}}_{\bm{k}, \alpha_1} \hat{\bar{b}}_{-\bm{k}, \alpha_2} \left( \hat{\bar{b}}_{\bm{0}, 0} \right)^{N - 2} \ket{0},
\end{equation}
which corresponds to the excitation of two quasiparticles specified by $(\bm{k}, \alpha_1)$ and $(-\bm{k}, \alpha_2)$ from the unperturbed steady state $\ket{\psi_\mathrm{st}^{(0)}}$.
For completeness, we give the expression of $\tilde{V}_{\alpha_1, \alpha_2, 0, 0} (\bm{k}, \bm{0}, \bm{0})$:
\begin{align}
    \tilde{V}_{\alpha_1, \alpha_2, 0, 0} (\bm{k}, \bm{0}, \bm{0}) & = \sum_{s_1, s_2} \left[ U D_0 (2 - \cos k_x - \cos k_y) - \frac{J \gamma_0}{2} s_1 s_2 \right] \nonumber \\
    & \ \ \ \ \ \ \ \ \ \ \ \times v_{\bm{k}, \alpha_1, s_1} v_{-\bm{k}, \alpha_2, s_2}.
    \label{eq_Vtilde_restricted}
\end{align}
We now have the analytical expression of the steady-state vector $\ket{\psi_\mathrm{st}}$ up to the first order of the interaction parameters $U$ and $J$ [Eqs.~\eqref{eq_psist0_explicit} and \eqref{eq_psist1_explicit}].
Since $\ket{\psi_\mathrm{st}}$ is equivalent to the steady-state distribution through $P_\mathrm{st} (\bm{n}) = {}_L \! \braket{\bm{n} | \psi_\mathrm{st}}$, we can calculate physical quantities using the obtained formula, as we explain in the next section.

\subsection{Correlation functions and structure factors}
\label{subsec_corr_strfac}

Using the expression of $\ket{\psi_\mathrm{st}} \simeq \ket{\psi_\mathrm{st}^{(0)}} + \ket{\psi_\mathrm{st}^{(1)}}$ obtained by perturbation theory [Eqs.~\eqref{eq_psist0_explicit} and \eqref{eq_psist1_explicit}], we expand the density structure factor $S_d (\bm{k})$ [Eq.~\eqref{eq_Sd_def}] up to the first order in $U$ and $J$:
\begin{equation}
    S_d (\bm{k}) \simeq S_d^{(0)} (\bm{k}) + S_d^{(1)} (\bm{k}).
    \label{eq_Sd_expansion}
\end{equation}

For $\bm{k} \neq \bm{0}$, the unperturbed structure factor $S_d^{(0)} (\bm{k})$ is given as
\begin{equation}
    S_d^{(0)} (\bm{k}) = \rho_0,
    \label{eq_Sd0}
\end{equation}
and the first-order correction $S_d^{(1)} (\bm{k})$ is obtained as
\begin{equation}
    S_d^{(1)} (\bm{k}) = -\frac{N (N - 1)}{L^4} \sum_{s, s'} \sum_{\alpha, \alpha'} \frac{u_{\bm{k}, \alpha, s} u_{-\bm{k}, \alpha', s'} \tilde{V}_{\alpha, \alpha', 0, 0} (\bm{k}, \bm{0}, \bm{0})}{\epsilon_{\bm{k}, \alpha} + \epsilon_{-\bm{k}, \alpha'}}.
    \label{eq_Sd1}
\end{equation}
See Eqs.~\eqref{eq_uk0}-\eqref{eq_epsilonk1} and \eqref{eq_Vtilde_restricted} for the expressions of $\bm{u}_{\bm{k}, \alpha}$, $\tilde{V}_{\alpha, \alpha', 0, 0} (\bm{k}, \bm{0}, \bm{0})$, and $\epsilon_{\bm{k}, \alpha}$.

For $\bm{k} = \bm{0}$, we can check $S_d^{(0)} (\bm{0}) = S_d^{(1)} (\bm{0}) = 0$, consistent with $S_d (\bm{0}) = 0$, which generally holds from the particle number conservation (i.e., $\sum_i n_i = N$).
Note that $S_d^{(0)} (\bm{k})$ [$= \rho_0 (1 - \delta_{\bm{k}, \bm{0}})$] leads to the unperturbed density correlation function:
\begin{equation}
    C_d^{(0)}(\bm{r}) := \frac{1}{L^2} \sum_{\bm{k}} e^{i \bm{k} \cdot \bm{r}} S_d(\bm{k}) = \rho_0 \delta_{\bm{r}, \bm{0}} - \frac{\rho_0}{L^2},
    \label{eq_Cd0}
\end{equation}
which includes a term $-\rho_0 / L^2$ due to the finite-size effect.
In Figs.~\ref{fig_repulsion}(e) and \ref{fig_alignment}(e), the two lines with light color represent the observed density correlation function after subtracting this term (i.e., $C_d (\bm{r}) + \rho_0 / L^2$).

Similarly, we expand the polarity structure factor as
\begin{equation}
    S_p (\bm{k}) \simeq S_p^{(0)} (\bm{k}) + S_p^{(1)} (\bm{k}),
    \label{eq_Sp_expansion}
\end{equation}
where
\begin{equation}
    S_p^{(0)} (\bm{k}) = \rho_0,
\end{equation}
and
\begin{equation}
    S_p^{(1)} (\bm{k}) = -\frac{N (N - 1)}{L^4} \sum_{s, s'} \sum_{\alpha, \alpha'} s s' \frac{u_{\bm{k}, \alpha, s} u_{-\bm{k}, \alpha', s'} \tilde{V}_{\alpha, \alpha', 0, 0} (\bm{k}, \bm{0}, \bm{0})}{\epsilon_{\bm{k}, \alpha} + \epsilon_{-\bm{k}, \alpha'}}.
    \label{eq_Sp1}
\end{equation}

We examine the long-wavelength behaviors of $S_d (\bm{k})$ and $S_p (\bm{k})$, which determine the long-distance behaviors of the correlation functions $C_d (\bm{r})$ and $C_p (\bm{r})$, respectively.
The terms with $(\alpha, \alpha') = (0, 0)$ in $S_d^{(1)} (\bm{k})$ [Eq.~\eqref{eq_Sd1}] and $S_p^{(1)} (\bm{k})$ [Eq.~\eqref{eq_Sp1}] can be non-analytic at $\bm{k} \to \bm{0}$, while the other terms with $(\alpha, \alpha') \neq (0, 0)$ are analytic.
Noticing that $\tilde{V}_{0, 0, 0, 0} (\bm{k}, \bm{0}, \bm{0}) \simeq U D_0 \bm{k}^2 - (J \varepsilon^2 {D_0}^2 / \gamma_0) {k_x}^2$ for $\bm{k} \simeq \bm{0}$, we can expand $S_d^{(1)} (\bm{k})$ and $S_p^{(1)} (\bm{k})$ for $\bm{k} \simeq \bm{0}$:
\begin{align}
    S_d^{(1)} (\bm{k}) & \simeq -\frac{N (N - 1)}{L^4} \frac{U \bm{k}^2 - J (\varepsilon^2 D_0 / \gamma_0) {k_x}^2}{\bm{k}^2 + (2 \varepsilon^2 D_0 / \gamma_0) {k_x}^2} + (\mathrm{a.t.})
    \label{eq_Sd1_approximate}
    \\
    S_p^{(1)} (\bm{k}) & \simeq \left( \frac{\varepsilon D_0}{\gamma_0} k_x \right)^2 S_d^{(1)} (\bm{k}) + (\mathrm{a.t.}) \nonumber \\
    & = -\frac{N (N - 1)}{L^4} \left( \frac{\varepsilon D_0}{\gamma_0} k_x \right)^2 \frac{U \bm{k}^2 - J (\varepsilon^2 D_0 / \gamma_0) {k_x}^2}{\bm{k}^2 + (2 \varepsilon^2 D_0 / \gamma_0) {k_x}^2} + (\mathrm{a.t.}),
    \label{eq_Sp1_approximate}
\end{align}
where $(\mathrm{a.t.})$ denotes the analytic terms derived from $(\alpha, \alpha') \neq (0, 0)$.
The first line in Eq.~\eqref{eq_Sp1_approximate} indicates the coupling of density and polarity at the level of first-order perturbation (see Sec.~\ref{sec_hydro} for the hydrodynamic counterpart).

\begin{figure}[t]
    \centering
    \includegraphics[scale=1]{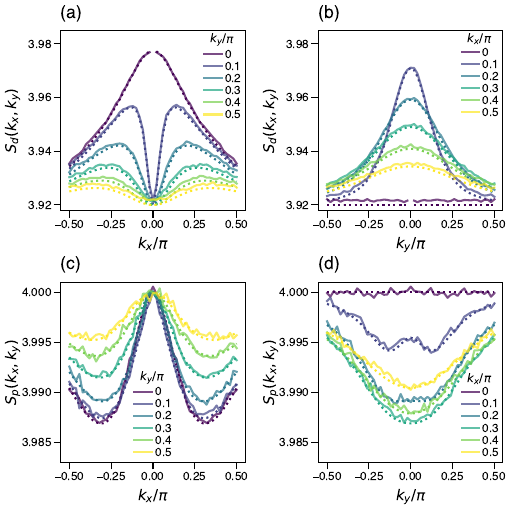}
    \caption{Singularity in structure factors induced by perturbative repulsion.
    (a,~b) Density structure factor $S_d (\bm{k})$ as a function of (a) $k_x$ and (b) $k_y$, for several fixed values of $k_y$ and $k_x$, respectively.
    (c,~d) Polarity structure factor $S_p (\bm{k})$, plotted in the same way as done in (a) and (b).
    The solid and dashed lines in each panel represent the values obtained from simulations and those predicted from perturbation theory.
    The predicted $S_d(\bm{k})$ and $S_p(\bm{k})$ are discontinuous and non-smooth at $\bm{k} \to \bm{0}$, respectively.
    The used parameters are $D_0 = 1$, $\varepsilon = 0.8$, $\gamma_0 = 0.5$, $U = 0.005$, $L = 100$, and $N = 4 \times 10^4$.}
    \label{fig_repulsion_compare}
\end{figure}

\begin{figure}[t]
    \centering
    \includegraphics[scale=1]{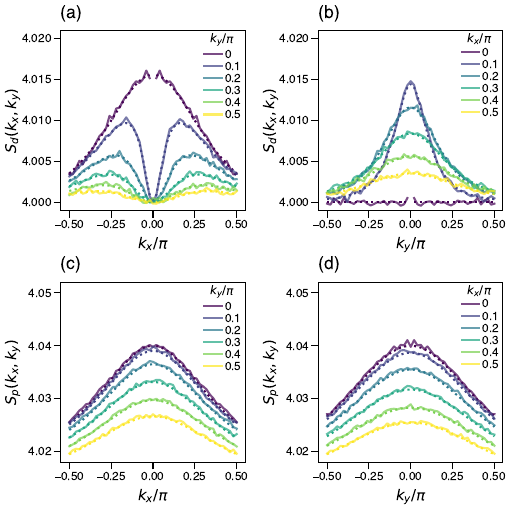}
    \caption{Singularity in structure factors induced by perturbative alignment.
    In the same way as shown in Fig.~\ref{fig_repulsion_compare}, we plot (a,~b) the density structure factor $S_d (\bm{k})$ and (c,~d) the polarity structure factor $S_p (\bm{k})$.
    The values obtained from simulations (solid lines) and those predicted from perturbation theory (dashed lines) show good agreement, suggesting singularities in $S_d (\bm{k})$ and $S_p (\bm{k})$ similarly to the case of repulsion (Fig.~\ref{fig_repulsion_compare}).
    The used parameters are $D_0 = 1$, $\varepsilon = 0.8$, $\gamma_0 = 2.0$, $J = 0.005$, $L = 100$, and $N = 4 \times 10^4$.}
    \label{fig_alignment_compare}
\end{figure}

From Eq.~\eqref{eq_Sd1_approximate}, we find that $S_d (\bm{k})$ is generically discontinuous at $\bm{k} \to \bm{0}$, which is characterized by the difference in the two limiting values:
\begin{align}
    & \lim_{k_x \to 0} S_d (k_x, 0) - \lim_{k_y \to 0} S_d (0, k_y) \nonumber\\
    & = \frac{N (N - 1)}{L^4} \frac{(2 U + J) \varepsilon^2 D_0 / \gamma_0}{1 + 2 \varepsilon^2 D_0 / \gamma_0},
    \label{eq_Sd_diff_analytic}
\end{align}
up to the first order in the repulsion $U$ ($\geq 0$) and alignment $J$ ($\geq 0$).
This formula suggests that the nonzero anisotropic self-propulsion $\varepsilon$ and interaction $U$ or $J$ are essential, but the interaction type (i.e., repulsion or alignment) is not essential, to the discontinuity of $S_d (\bm{k})$.
In addition, the prefactor $N (N - 1)$ suggests that the discontinuity appears even at the two-particle level (i.e., $N = 2$).
As in the case of the hydrodynamic model (see Sec.~\ref{sec_hydro}), from $C_d (\bm{r}) = L^{-2} \sum_{\bm{k}} e^{i \bm{k} \cdot \bm{r}} S_d (\bm{k}) \simeq \int d^2 \bm{k} e^{i \bm{k} \cdot \bm{r}} S_d (\bm{k})$, we find that the density correlation function follows a power law for $|\bm{r}| \to \infty$ reflecting the discontinuity of $S_d (\bm{k})$:
\begin{equation}
    \begin{split}
        C_d (x, 0) & \sim x^{-2} \\
        C_d (0, y) & \sim y^{-2}.
    \end{split}
    \label{eq_Cd_power_law}
\end{equation}
Note that the power-law density correlation in two-particle systems has also been found in externally driven systems~\cite{Sasa2005,Lefevere2005}.

Focusing on Eq.~\eqref{eq_Sp1_approximate}, we see that $S_p (\bm{k})$ is continuous but can be non-smooth at $\bm{k} \to \bm{0}$, which is characterized by
\begin{align}
    & \lim_{k_x \to 0} \frac{\partial^2 S_p (\bm{k})}{{\partial k_y}^2} \Bigg|_{(k_x, 0)} - \lim_{k_y \to 0} \frac{\partial^2 S_p (\bm{k})}{{\partial k_y}^2} \Bigg|_{(0, k_y)} \nonumber \\
    & = -\frac{N (N - 1)}{L^4} \frac{2 (2 U + J) \varepsilon^4 (D_0 / \gamma_0)^3}{(1 + 2 \varepsilon^2 D_0 / \gamma_0)^2},
    \label{eq_Sp_diff_analytic}
\end{align}
up to the first order in $U$ and $J$.
The condition for the non-smoothness of $S_p (\bm{k})$ is the same as that for the discontinuity of $S_d (\bm{k})$, i.e., nonzero anisotropic self-propulsion and interaction (regardless of repulsion or alignment).
From $C_p (\bm{r}) = L^{-2} \sum_{\bm{k}} e^{i \bm{k} \cdot \bm{r}} S_p (\bm{k}) \simeq \int d^2 \bm{k} e^{i \bm{k} \cdot \bm{r}} S_p (\bm{k})$, we find that the power-law correlation of polarity appears for $|\bm{r}| \to \infty$ reflecting the non-smoothness of $S_p (\bm{k})$:
\begin{equation}
    \begin{split}
        C_p (x, 0) & \sim x^{-4} \\
        C_p (0, y) & \sim y^{-4}.
    \end{split}
    \label{eq_Cp_power_law}
\end{equation}

\subsection{Comparison with numerical results}

In the following, we confirm that the obtained formulas [Eqs.~\eqref{eq_Sd_expansion}-\eqref{eq_Sd1} and Eqs.~\eqref{eq_Sp_expansion}-\eqref{eq_Sp1}] quantitatively reproduce the numerical results when $U$ and $J$ are small.

We first consider a purely repulsive interaction (i.e., $U > 0$ and $J = 0$).
The parameters are set as $D_0 = 1$, $\varepsilon = 0.8$, $\gamma_0 = 0.5$, $U = 0.005$, $L = 100$, and $N = 4 \times 10^4$ (corresponding to $\rho_0 = 4$).
In Figs.~\ref{fig_repulsion_compare}(a) and (b), we compare the numerically obtained density structure factor $S_d (\bm{k})$ (solid lines) and the analytical results based on Eqs.~\eqref{eq_Sd_expansion}-\eqref{eq_Sd1} (dotted lines).
Both the (a) $k_x$ dependence and (b) $k_y$ dependence of $S_d (\bm{k})$ are reproduced well by the analytical formulas, especially when $S_d (\bm{k})$ is close to the unperturbed value $S_d^{(0)} (\bm{k}) = \rho_0 = 4$.
In Figs.~\ref{fig_repulsion_compare}(c) and (d), we compare the numerically obtained polarity structure factor $S_p (\bm{k})$ (solid lines) and the analytical results based on Eqs.~\eqref{eq_Sp_expansion}-\eqref{eq_Sp1} (dotted lines), which also show good agreement.

We next consider a purely aligning interaction (i.e., $U = 0$ and $J > 0$).
The parameters are $D_0 = 1$, $\varepsilon = 0.8$, $\gamma_0 = 2$, $J = 0.005$, $L = 100$, and $N = 4 \times 10^4$ (corresponding to $\rho_0 = 4$).
In Fig.~\ref{fig_alignment_compare}, the figures are placed in the same order as Fig.~\ref{fig_repulsion_compare}.
In all figures, the numerical results agree with the analytical predictions with high accuracy.

\section{Discussion and summary}

In this paper, we have studied the correlation properties in the homogeneous disordered state of particle systems with uniaxial self-propulsion.
Starting with hydrodynamic arguments, we have proposed that such systems should generically show power-law density correlation and polarity correlation with exponent $-2$ and $-4$, respectively.
Performing simulations of a lattice gas model with uniaxial self-propulsion and repulsion or alignment between particles, we have found that the predicted power-law correlations indeed appear regardless of the interaction type.
Further, using the Doi-Peliti method and perturbation theory, we have mapped the model to a two-component boson system and analyzed the effects of interaction as quasiparticle excitations.
We have analytically obtained the formulas for the density and polarity structure factors, which have singularities that lead to the power-law decay of correlation functions, even in the first order of the interaction strength.
Both the hydrodynamic and microscopic formulas [Eqs.~\eqref{eq_w_adiabatic_approx} and \eqref{eq_Sp1_approximate}] suggest that the coupling of density and polarity is essential to the power-law correlation of polarity.

We have focused on the homogeneous disordered state, which generically appears when the interaction strength is weak compared to the noise effect [Figs.~\ref{fig_phase}(a) and (c)].
When the interaction is strong, phase separation or long-range order can appear through phase transitions such as MIPS and flocking [Figs.~\ref{fig_phase}(b) and (d)].
It is interesting to examine whether the power-law correlation for density fluctuation or polarity fluctuation still appears in such inhomogeneous or ordered states.

According to our results, in anisotropic biological systems, we may observe the power-law correlations of density or polarity as a universal nonequilibrium collective phenomenon, regardless of the detail of interactions.
For example, spatial anisotropy can be introduced as an anisotropic interaction between a cell population and orientationally aligned substrate~\cite{Luo2023,Gu2023,Skillin2023}.
We note that observation of the power law using a few samples can be hard, given that the correlation functions decay relatively quickly since the exponents are $-2$ and $-4$ for density and polarity, respectively [i.e., $C_d (r) \sim r^{-2}$ and $C_p (r) \sim r^{-4}$].
Nevertheless, it will be interesting to probe nonequilibrium properties inherent in biological systems by adding spatial anisotropy.

\begin{acknowledgments}
We thank Ryusuke Hamazaki, Kyogo Kawaguchi, Hiroshi Noguchi, Yuta Sekino, Kazuaki Takasan, Takeru Yokota for fruitful discussions.
The computations in this study were performed using the facilities of the Supercomputer Center at the Institute for Solid State Physics, the University of Tokyo.
This work was supported by JSPS KAKENHI Grant Numbers JP20K14435 (to K.A.) and JP22K13978 (to H.N.).
\end{acknowledgments}

\appendix

\begin{figure}[t]
    \centering
    \includegraphics[scale=1]{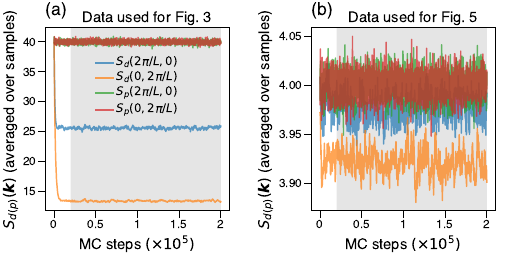}
    \caption{Relaxation dynamics of structure factors for the model with repulsion.
    We plot the time evolution of the long-wavelength components of the structure factors $S_d(\bm{k})$ and $S_p(\bm{k})$ averaged over independent samples for the parameter sets used in (a) Fig.~\ref{fig_repulsion} and (b) Fig.~\ref{fig_repulsion_compare}.
    The gray area in each panel suggests the time points used for averaging.}
    \label{fig_repulsion_timedep}
\end{figure}

\begin{figure}[t]
    \centering
    \includegraphics[scale=1]{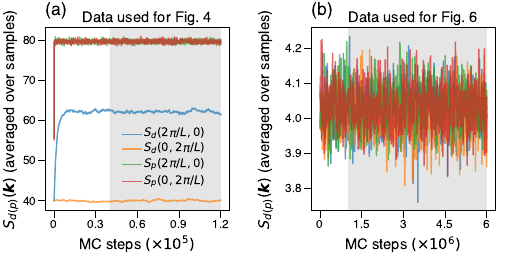}
    \caption{Relaxation dynamics of structure factors for the model with alignment.
    We plot the time evolution of the long-wavelength components of the structure factors $S_d(\bm{k})$ and $S_p(\bm{k})$ averaged over independent samples for the parameter sets used in (a) Fig.~\ref{fig_alignment} and (b) Fig.~\ref{fig_alignment_compare}.
    The gray area in each panel suggests the time points used for averaging.}
    \label{fig_alignment_timedep}
\end{figure}

\section{Numerical simulation of lattice gas model}
\label{app_lattice_gas_model}

For the simulation of the lattice gas model explained in Sec.~\ref{subsec_lattice_gas_model}, we discretize the time with interval $\Delta t$.
For each sample, we generate the initial configuration of $N$ particles by placing each particle on a randomly chosen site with a randomly chosen spin.
In a single Monte Carlo (MC) step, we update the configuration as follows.
\begin{enumerate}[label=(\arabic*)]
    \item We randomly choose a particle.
    \item The chosen particle (with spin $s$ at site $i$) (i) hops by one site or (ii) flips the spin with the following probabilities.
    \begin{enumerate}[label=(\roman*)]
        \item The particle hops one site right with probability $2 (1 + s \varepsilon) D_0 \Delta t / [1 + e^{U (\Delta n_{i \to j} + 1)}]$, left with probability $2 (1 - s \varepsilon) D_0 \Delta t / [1 + e^{U (\Delta n_{i \to j} + 1)}]$, or up or down with probability $2 D_0 \Delta t / [1 + e^{U (\Delta n_{i \to j} + 1)}]$ for each, where $\Delta n_{i \to j} := n_j - n_i$ is the difference in the local density between the target site $j$ and the departure site $i$.
        \item The spin of the particle flips to $-s$ with probability $2 \gamma_0 \Delta t / [1 + e^{J (s m_i - 1)}]$, where $m_i = \sum_{s'} s' n_{i, s'}$ is the local polarity at site $i$.
    \end{enumerate}
    \item We repeat the procedures (1) and (2) $N$ times and increment time by $\Delta t$.
\end{enumerate}

We set $\Delta t = 1 / (8 D_0 + \gamma_0)$ for the simulations of the model with purely repulsive interaction (i.e., $U > 0$ and $J = 0$) (Figs.~\ref{fig_repulsion} and \ref{fig_repulsion_compare}) and $\Delta t = 1 / (4 D_0 + 2 \gamma_0)$ for the simulations of the model with purely aligning interaction (i.e., $U = 0$ and $J > 0$) (Figs.~\ref{fig_alignment} and \ref{fig_alignment_compare}).

\section{Numerical sampling of configurations}
\label{app_sampling}

In the lattice gas model simulations, we checked the relaxation of the long-wavelength components of the structure factors, $S_d (\bm{k})$ and $S_p (\bm{k})$ for $\bm{k} = (2\pi / L, 0)$ and $\bm{k} = (0, 2\pi / L)$, which is slower than that of the short-wavelength components.
In Figs.~\ref{fig_repulsion_timedep}(a) and (b), we plot the relaxation dynamics of these quantities averaged over independent samples with the parameter sets used for Figs.~\ref{fig_repulsion} and \ref{fig_repulsion_compare} (purely repulsive interaction), respectively.
The gray area in each panel suggests the time points used for averaging.
Specifically, we took averages over $18432$ independent samples and $1800$ time points (taken every 100 MC steps after relaxation with $2 \times 10^4$ MC steps) to obtain the structure factors and correlation functions for Fig.~\ref{fig_repulsion}; $92160$ independent samples and $1800$ time points (taken every 100 MC steps after relaxation with $2 \times 10^4$ MC steps) for Fig.~\ref{fig_repulsion_compare}. 

In Figs.~\ref{fig_alignment_timedep}(a) and (b), we plot the corresponding relaxation dynamics with the parameter sets used for Figs.~\ref{fig_alignment} and \ref{fig_alignment_compare} (purely aligning interaction), respectively.
We took averages over $44800$ independent samples and $800$ time points (taken every 100 MC steps after relaxation with $4 \times 10^4$ MC steps) for Figs.~\ref{fig_alignment}; $4096$ independent samples and $50000$ time points (taken every 100 MC steps after relaxation with $1 \times 10^6$ MC steps) for Figs.~\ref{fig_alignment_compare}.


\begin{thebibliography}{50}%
\makeatletter
\providecommand \@ifxundefined [1]{%
 \@ifx{#1\undefined}
}%
\providecommand \@ifnum [1]{%
 \ifnum #1\expandafter \@firstoftwo
 \else \expandafter \@secondoftwo
 \fi
}%
\providecommand \@ifx [1]{%
 \ifx #1\expandafter \@firstoftwo
 \else \expandafter \@secondoftwo
 \fi
}%
\providecommand \natexlab [1]{#1}%
\providecommand \enquote  [1]{``#1''}%
\providecommand \bibnamefont  [1]{#1}%
\providecommand \bibfnamefont [1]{#1}%
\providecommand \citenamefont [1]{#1}%
\providecommand \href@noop [0]{\@secondoftwo}%
\providecommand \href [0]{\begingroup \@sanitize@url \@href}%
\providecommand \@href[1]{\@@startlink{#1}\@@href}%
\providecommand \@@href[1]{\endgroup#1\@@endlink}%
\providecommand \@sanitize@url [0]{\catcode `\\12\catcode `\$12\catcode
  `\&12\catcode `\#12\catcode `\^12\catcode `\_12\catcode `\%12\relax}%
\providecommand \@@startlink[1]{}%
\providecommand \@@endlink[0]{}%
\providecommand \url  [0]{\begingroup\@sanitize@url \@url }%
\providecommand \@url [1]{\endgroup\@href {#1}{\urlprefix }}%
\providecommand \urlprefix  [0]{URL }%
\providecommand \Eprint [0]{\href }%
\providecommand \doibase [0]{https://doi.org/}%
\providecommand \selectlanguage [0]{\@gobble}%
\providecommand \bibinfo  [0]{\@secondoftwo}%
\providecommand \bibfield  [0]{\@secondoftwo}%
\providecommand \translation [1]{[#1]}%
\providecommand \BibitemOpen [0]{}%
\providecommand \bibitemStop [0]{}%
\providecommand \bibitemNoStop [0]{.\EOS\space}%
\providecommand \EOS [0]{\spacefactor3000\relax}%
\providecommand \BibitemShut  [1]{\csname bibitem#1\endcsname}%
\let\auto@bib@innerbib\@empty
\bibitem [{\citenamefont {Marchetti}\ \emph {et~al.}(2013)\citenamefont
  {Marchetti}, \citenamefont {Joanny}, \citenamefont {Ramaswamy}, \citenamefont
  {Liverpool}, \citenamefont {Prost}, \citenamefont {Rao},\ and\ \citenamefont
  {Simha}}]{Marchetti2013}%
  \BibitemOpen
  \bibfield  {author} {\bibinfo {author} {\bibfnamefont {M.~C.}\ \bibnamefont
  {Marchetti}}, \bibinfo {author} {\bibfnamefont {J.~F.}\ \bibnamefont
  {Joanny}}, \bibinfo {author} {\bibfnamefont {S.}~\bibnamefont {Ramaswamy}},
  \bibinfo {author} {\bibfnamefont {T.~B.}\ \bibnamefont {Liverpool}}, \bibinfo
  {author} {\bibfnamefont {J.}~\bibnamefont {Prost}}, \bibinfo {author}
  {\bibfnamefont {M.}~\bibnamefont {Rao}},\ and\ \bibinfo {author}
  {\bibfnamefont {R.~A.}\ \bibnamefont {Simha}},\ }\bibfield  {title} {\bibinfo
  {title} {Hydrodynamics of soft active matter},\ }\href
  {https://doi.org/10.1103/RevModPhys.85.1143} {\bibfield  {journal} {\bibinfo
  {journal} {Rev. Mod. Phys.}\ }\textbf {\bibinfo {volume} {85}},\ \bibinfo
  {pages} {1143} (\bibinfo {year} {2013})}\BibitemShut {NoStop}%
\bibitem [{\citenamefont {Needleman}\ and\ \citenamefont
  {Dogic}(2017)}]{Needleman2017}%
  \BibitemOpen
  \bibfield  {author} {\bibinfo {author} {\bibfnamefont {D.}~\bibnamefont
  {Needleman}}\ and\ \bibinfo {author} {\bibfnamefont {Z.}~\bibnamefont
  {Dogic}},\ }\bibfield  {title} {\bibinfo {title} {Active matter at the
  interface between materials science and cell biology},\ }\href
  {https://doi.org/10.1038/natrevmats.2017.48} {\bibfield  {journal} {\bibinfo
  {journal} {Nat. Rev. Mater.}\ }\textbf {\bibinfo {volume} {2}},\ \bibinfo
  {pages} {1} (\bibinfo {year} {2017})}\BibitemShut {NoStop}%
\bibitem [{\citenamefont {Gompper}\ \emph {et~al.}(2020)\citenamefont
  {Gompper}, \citenamefont {Winkler}, \citenamefont {Speck}, \citenamefont
  {Solon}, \citenamefont {Nardini}, \citenamefont {Peruani}, \citenamefont
  {L{\"o}wen}, \citenamefont {Golestanian}, \citenamefont {Kaupp},
  \citenamefont {Alvarez}, \citenamefont {Ki{\o}rboe}, \citenamefont {Lauga},
  \citenamefont {Poon}, \citenamefont {DeSimone}, \citenamefont
  {Mui{\~n}os-Landin}, \citenamefont {Fischer}, \citenamefont {S{\"o}ker},
  \citenamefont {Cichos}, \citenamefont {Kapral}, \citenamefont {Gaspard},
  \citenamefont {Ripoll}, \citenamefont {Sagues}, \citenamefont
  {Doostmohammadi}, \citenamefont {Yeomans}, \citenamefont {Aranson},
  \citenamefont {Bechinger}, \citenamefont {Stark}, \citenamefont {Hemelrijk},
  \citenamefont {Nedelec}, \citenamefont {Sarkar}, \citenamefont {Aryaksama},
  \citenamefont {Lacroix}, \citenamefont {Duclos}, \citenamefont {Yashunsky},
  \citenamefont {Silberzan}, \citenamefont {Arroyo},\ and\ \citenamefont
  {Kale}}]{Gompper2020}%
  \BibitemOpen
  \bibfield  {author} {\bibinfo {author} {\bibfnamefont {G.}~\bibnamefont
  {Gompper}}, \bibinfo {author} {\bibfnamefont {R.~G.}\ \bibnamefont
  {Winkler}}, \bibinfo {author} {\bibfnamefont {T.}~\bibnamefont {Speck}},
  \bibinfo {author} {\bibfnamefont {A.}~\bibnamefont {Solon}}, \bibinfo
  {author} {\bibfnamefont {C.}~\bibnamefont {Nardini}}, \bibinfo {author}
  {\bibfnamefont {F.}~\bibnamefont {Peruani}}, \bibinfo {author} {\bibfnamefont
  {H.}~\bibnamefont {L{\"o}wen}}, \bibinfo {author} {\bibfnamefont
  {R.}~\bibnamefont {Golestanian}}, \bibinfo {author} {\bibfnamefont {U.~B.}\
  \bibnamefont {Kaupp}}, \bibinfo {author} {\bibfnamefont {L.}~\bibnamefont
  {Alvarez}}, \bibinfo {author} {\bibfnamefont {T.}~\bibnamefont {Ki{\o}rboe}},
  \bibinfo {author} {\bibfnamefont {E.}~\bibnamefont {Lauga}}, \bibinfo
  {author} {\bibfnamefont {W.~C.~K.}\ \bibnamefont {Poon}}, \bibinfo {author}
  {\bibfnamefont {A.}~\bibnamefont {DeSimone}}, \bibinfo {author}
  {\bibfnamefont {S.}~\bibnamefont {Mui{\~n}os-Landin}}, \bibinfo {author}
  {\bibfnamefont {A.}~\bibnamefont {Fischer}}, \bibinfo {author} {\bibfnamefont
  {N.~A.}\ \bibnamefont {S{\"o}ker}}, \bibinfo {author} {\bibfnamefont
  {F.}~\bibnamefont {Cichos}}, \bibinfo {author} {\bibfnamefont
  {R.}~\bibnamefont {Kapral}}, \bibinfo {author} {\bibfnamefont
  {P.}~\bibnamefont {Gaspard}}, \bibinfo {author} {\bibfnamefont
  {M.}~\bibnamefont {Ripoll}}, \bibinfo {author} {\bibfnamefont
  {F.}~\bibnamefont {Sagues}}, \bibinfo {author} {\bibfnamefont
  {A.}~\bibnamefont {Doostmohammadi}}, \bibinfo {author} {\bibfnamefont
  {J.~M.}\ \bibnamefont {Yeomans}}, \bibinfo {author} {\bibfnamefont {I.~S.}\
  \bibnamefont {Aranson}}, \bibinfo {author} {\bibfnamefont {C.}~\bibnamefont
  {Bechinger}}, \bibinfo {author} {\bibfnamefont {H.}~\bibnamefont {Stark}},
  \bibinfo {author} {\bibfnamefont {C.~K.}\ \bibnamefont {Hemelrijk}}, \bibinfo
  {author} {\bibfnamefont {F.~J.}\ \bibnamefont {Nedelec}}, \bibinfo {author}
  {\bibfnamefont {T.}~\bibnamefont {Sarkar}}, \bibinfo {author} {\bibfnamefont
  {T.}~\bibnamefont {Aryaksama}}, \bibinfo {author} {\bibfnamefont
  {M.}~\bibnamefont {Lacroix}}, \bibinfo {author} {\bibfnamefont
  {G.}~\bibnamefont {Duclos}}, \bibinfo {author} {\bibfnamefont
  {V.}~\bibnamefont {Yashunsky}}, \bibinfo {author} {\bibfnamefont
  {P.}~\bibnamefont {Silberzan}}, \bibinfo {author} {\bibfnamefont
  {M.}~\bibnamefont {Arroyo}},\ and\ \bibinfo {author} {\bibfnamefont
  {S.}~\bibnamefont {Kale}},\ }\bibfield  {title} {\bibinfo {title} {The 2020
  motile active matter roadmap},\ }\href
  {https://doi.org/10.1088/1361-648X/ab6348} {\bibfield  {journal} {\bibinfo
  {journal} {J. Phys. Condens. Matter}\ }\textbf {\bibinfo {volume} {32}},\
  \bibinfo {pages} {193001} (\bibinfo {year} {2020})}\BibitemShut {NoStop}%
\bibitem [{\citenamefont {Vicsek}\ \emph {et~al.}(1995)\citenamefont {Vicsek},
  \citenamefont {Czir{\'o}k}, \citenamefont {Ben-Jacob}, \citenamefont
  {Cohen},\ and\ \citenamefont {Shochet}}]{Vicsek1995}%
  \BibitemOpen
  \bibfield  {author} {\bibinfo {author} {\bibfnamefont {T.}~\bibnamefont
  {Vicsek}}, \bibinfo {author} {\bibfnamefont {A.}~\bibnamefont {Czir{\'o}k}},
  \bibinfo {author} {\bibfnamefont {E.}~\bibnamefont {Ben-Jacob}}, \bibinfo
  {author} {\bibfnamefont {I.}~\bibnamefont {Cohen}, \bibfnamefont {I}},\ and\
  \bibinfo {author} {\bibfnamefont {O.}~\bibnamefont {Shochet}},\ }\bibfield
  {title} {\bibinfo {title} {Novel type of phase transition in a system of
  self-driven particles},\ }\href {https://doi.org/10.1103/PhysRevLett.75.1226}
  {\bibfield  {journal} {\bibinfo  {journal} {Phys. Rev. Lett.}\ }\textbf
  {\bibinfo {volume} {75}},\ \bibinfo {pages} {1226} (\bibinfo {year}
  {1995})}\BibitemShut {NoStop}%
\bibitem [{\citenamefont {Toner}\ and\ \citenamefont {Tu}(1995)}]{Toner1995}%
  \BibitemOpen
  \bibfield  {author} {\bibinfo {author} {\bibfnamefont {J.}~\bibnamefont
  {Toner}}\ and\ \bibinfo {author} {\bibfnamefont {Y.}~\bibnamefont {Tu}},\
  }\bibfield  {title} {\bibinfo {title} {{Long-Range} order in a
  {Two-Dimensional} dynamical {XY} model: How birds fly together},\ }\href
  {https://doi.org/10.1103/PhysRevLett.75.4326} {\bibfield  {journal} {\bibinfo
   {journal} {Phys. Rev. Lett.}\ }\textbf {\bibinfo {volume} {75}},\ \bibinfo
  {pages} {4326} (\bibinfo {year} {1995})}\BibitemShut {NoStop}%
\bibitem [{\citenamefont {Toner}(2012)}]{Toner2012}%
  \BibitemOpen
  \bibfield  {author} {\bibinfo {author} {\bibfnamefont {J.}~\bibnamefont
  {Toner}},\ }\bibfield  {title} {\bibinfo {title} {Reanalysis of the
  hydrodynamic theory of fluid, polar-ordered flocks},\ }\href
  {https://doi.org/10.1103/PhysRevE.86.031918} {\bibfield  {journal} {\bibinfo
  {journal} {Phys. Rev. E}\ }\textbf {\bibinfo {volume} {86}},\ \bibinfo
  {pages} {031918} (\bibinfo {year} {2012})}\BibitemShut {NoStop}%
\bibitem [{\citenamefont {Chat{\'e}}(2020)}]{Chate2020}%
  \BibitemOpen
  \bibfield  {author} {\bibinfo {author} {\bibfnamefont {H.}~\bibnamefont
  {Chat{\'e}}},\ }\bibfield  {title} {\bibinfo {title} {Dry aligning dilute
  active matter},\ }\href
  {https://doi.org/10.1146/annurev-conmatphys-031119-050752} {\bibfield
  {journal} {\bibinfo  {journal} {Annu. Rev. Condens. Matter Phys.}\ }\textbf
  {\bibinfo {volume} {11}},\ \bibinfo {pages} {189} (\bibinfo {year}
  {2020})}\BibitemShut {NoStop}%
\bibitem [{\citenamefont {Cates}\ and\ \citenamefont
  {Tailleur}(2015)}]{Cates2015}%
  \BibitemOpen
  \bibfield  {author} {\bibinfo {author} {\bibfnamefont {M.~E.}\ \bibnamefont
  {Cates}}\ and\ \bibinfo {author} {\bibfnamefont {J.}~\bibnamefont
  {Tailleur}},\ }\bibfield  {title} {\bibinfo {title} {{Motility-Induced} phase
  separation},\ }\href
  {https://doi.org/10.1146/annurev-conmatphys-031214-014710} {\bibfield
  {journal} {\bibinfo  {journal} {Annu. Rev. Condens. Matter Phys.}\ }\textbf
  {\bibinfo {volume} {6}},\ \bibinfo {pages} {219} (\bibinfo {year}
  {2015})}\BibitemShut {NoStop}%
\bibitem [{\citenamefont {Solon}\ and\ \citenamefont
  {Tailleur}(2013)}]{Solon2013}%
  \BibitemOpen
  \bibfield  {author} {\bibinfo {author} {\bibfnamefont {A.~P.}\ \bibnamefont
  {Solon}}\ and\ \bibinfo {author} {\bibfnamefont {J.}~\bibnamefont
  {Tailleur}},\ }\bibfield  {title} {\bibinfo {title} {Revisiting the flocking
  transition using active spins},\ }\href
  {https://doi.org/10.1103/PhysRevLett.111.078101} {\bibfield  {journal}
  {\bibinfo  {journal} {Phys. Rev. Lett.}\ }\textbf {\bibinfo {volume} {111}},\
  \bibinfo {pages} {078101} (\bibinfo {year} {2013})}\BibitemShut {NoStop}%
\bibitem [{\citenamefont {Solon}\ and\ \citenamefont
  {Tailleur}(2015)}]{Solon2015a}%
  \BibitemOpen
  \bibfield  {author} {\bibinfo {author} {\bibfnamefont {A.~P.}\ \bibnamefont
  {Solon}}\ and\ \bibinfo {author} {\bibfnamefont {J.}~\bibnamefont
  {Tailleur}},\ }\bibfield  {title} {\bibinfo {title} {Flocking with discrete
  symmetry: The two-dimensional active ising model},\ }\href
  {https://doi.org/10.1103/PhysRevE.92.042119} {\bibfield  {journal} {\bibinfo
  {journal} {Phys. Rev. E}\ }\textbf {\bibinfo {volume} {92}},\ \bibinfo
  {pages} {042119} (\bibinfo {year} {2015})}\BibitemShut {NoStop}%
\bibitem [{\citenamefont {Solon}\ \emph {et~al.}(2015)\citenamefont {Solon},
  \citenamefont {Chat{\'e}},\ and\ \citenamefont {Tailleur}}]{Solon2015b}%
  \BibitemOpen
  \bibfield  {author} {\bibinfo {author} {\bibfnamefont {A.~P.}\ \bibnamefont
  {Solon}}, \bibinfo {author} {\bibfnamefont {H.}~\bibnamefont {Chat{\'e}}},\
  and\ \bibinfo {author} {\bibfnamefont {J.}~\bibnamefont {Tailleur}},\
  }\bibfield  {title} {\bibinfo {title} {From phase to microphase separation in
  flocking models: the essential role of nonequilibrium fluctuations},\ }\href
  {https://doi.org/10.1103/PhysRevLett.114.068101} {\bibfield  {journal}
  {\bibinfo  {journal} {Phys. Rev. Lett.}\ }\textbf {\bibinfo {volume} {114}},\
  \bibinfo {pages} {068101} (\bibinfo {year} {2015})}\BibitemShut {NoStop}%
\bibitem [{\citenamefont {Ngo}\ \emph {et~al.}(2014)\citenamefont {Ngo},
  \citenamefont {Peshkov}, \citenamefont {Aranson}, \citenamefont {Bertin},
  \citenamefont {Ginelli},\ and\ \citenamefont {Chat{\'e}}}]{Ngo2014}%
  \BibitemOpen
  \bibfield  {author} {\bibinfo {author} {\bibfnamefont {S.}~\bibnamefont
  {Ngo}}, \bibinfo {author} {\bibfnamefont {A.}~\bibnamefont {Peshkov}},
  \bibinfo {author} {\bibfnamefont {I.~S.}\ \bibnamefont {Aranson}}, \bibinfo
  {author} {\bibfnamefont {E.}~\bibnamefont {Bertin}}, \bibinfo {author}
  {\bibfnamefont {F.}~\bibnamefont {Ginelli}},\ and\ \bibinfo {author}
  {\bibfnamefont {H.}~\bibnamefont {Chat{\'e}}},\ }\bibfield  {title} {\bibinfo
  {title} {Large-scale chaos and fluctuations in active nematics},\ }\href
  {https://doi.org/10.1103/PhysRevLett.113.038302} {\bibfield  {journal}
  {\bibinfo  {journal} {Phys. Rev. Lett.}\ }\textbf {\bibinfo {volume} {113}},\
  \bibinfo {pages} {038302} (\bibinfo {year} {2014})}\BibitemShut {NoStop}%
\bibitem [{\citenamefont {Garrido}\ \emph {et~al.}(1990)\citenamefont
  {Garrido}, \citenamefont {Lebowitz}, \citenamefont {Maes},\ and\
  \citenamefont {Spohn}}]{Garrido1990}%
  \BibitemOpen
  \bibfield  {author} {\bibinfo {author} {\bibfnamefont {P.~L.}\ \bibnamefont
  {Garrido}}, \bibinfo {author} {\bibfnamefont {J.~L.}\ \bibnamefont
  {Lebowitz}}, \bibinfo {author} {\bibfnamefont {C.}~\bibnamefont {Maes}},\
  and\ \bibinfo {author} {\bibfnamefont {H.}~\bibnamefont {Spohn}},\ }\bibfield
   {title} {\bibinfo {title} {Long-range correlations for conservative
  dynamics},\ }\href {https://doi.org/10.1103/physreva.42.1954} {\bibfield
  {journal} {\bibinfo  {journal} {Phys. Rev. A}\ }\textbf {\bibinfo {volume}
  {42}},\ \bibinfo {pages} {1954} (\bibinfo {year} {1990})}\BibitemShut
  {NoStop}%
\bibitem [{\citenamefont {Dorfman}\ \emph {et~al.}(1994)\citenamefont
  {Dorfman}, \citenamefont {Kirkpatrick},\ and\ \citenamefont
  {Sengers}}]{Dorfman1994}%
  \BibitemOpen
  \bibfield  {author} {\bibinfo {author} {\bibfnamefont {J.~R.}\ \bibnamefont
  {Dorfman}}, \bibinfo {author} {\bibfnamefont {T.~R.}\ \bibnamefont
  {Kirkpatrick}},\ and\ \bibinfo {author} {\bibfnamefont {J.~V.}\ \bibnamefont
  {Sengers}},\ }\bibfield  {title} {\bibinfo {title} {Generic {Long-Range}
  correlations in molecular fluids},\ }\href
  {https://doi.org/10.1146/annurev.pc.45.100194.001241} {\bibfield  {journal}
  {\bibinfo  {journal} {Annu. Rev. Phys. Chem.}\ }\textbf {\bibinfo {volume}
  {45}},\ \bibinfo {pages} {213} (\bibinfo {year} {1994})}\BibitemShut
  {NoStop}%
\bibitem [{\citenamefont {Schmittmann}\ and\ \citenamefont
  {Zia}(1995)}]{Schmittmann1995}%
  \BibitemOpen
  \bibfield  {author} {\bibinfo {author} {\bibfnamefont {B.}~\bibnamefont
  {Schmittmann}}\ and\ \bibinfo {author} {\bibfnamefont {R.~K.~P.}\
  \bibnamefont {Zia}},\ }\bibfield  {title} {\bibinfo {title} {Statistical
  mechanics of driven diffusive systems},\ }in\ \href@noop {} {\emph {\bibinfo
  {booktitle} {Phase Transitions and Critical Phenomena}}},\ Vol.~\bibinfo
  {volume} {17}\ (\bibinfo  {publisher} {Academic Press},\ \bibinfo {year}
  {1995})\BibitemShut {NoStop}%
\bibitem [{\citenamefont {Schmittmann}\ and\ \citenamefont
  {Zia}(1998)}]{Schmittmann1998}%
  \BibitemOpen
  \bibfield  {author} {\bibinfo {author} {\bibfnamefont {B.}~\bibnamefont
  {Schmittmann}}\ and\ \bibinfo {author} {\bibfnamefont {R.~K.~P.}\
  \bibnamefont {Zia}},\ }\bibfield  {title} {\bibinfo {title} {Driven diffusive
  systems. an introduction and recent developments},\ }\href
  {https://doi.org/10.1016/S0370-1573(98)00005-2} {\bibfield  {journal}
  {\bibinfo  {journal} {Phys. Rep.}\ }\textbf {\bibinfo {volume} {301}},\
  \bibinfo {pages} {45} (\bibinfo {year} {1998})}\BibitemShut {NoStop}%
\bibitem [{\citenamefont {Adachi}\ \emph {et~al.}(2022)\citenamefont {Adachi},
  \citenamefont {Takasan},\ and\ \citenamefont {Kawaguchi}}]{Adachi2022}%
  \BibitemOpen
  \bibfield  {author} {\bibinfo {author} {\bibfnamefont {K.}~\bibnamefont
  {Adachi}}, \bibinfo {author} {\bibfnamefont {K.}~\bibnamefont {Takasan}},\
  and\ \bibinfo {author} {\bibfnamefont {K.}~\bibnamefont {Kawaguchi}},\
  }\bibfield  {title} {\bibinfo {title} {Activity-induced phase transition in a
  quantum many-body system},\ }\href
  {https://doi.org/10.1103/PhysRevResearch.4.013194} {\bibfield  {journal}
  {\bibinfo  {journal} {Phys. Rev. Res.}\ }\textbf {\bibinfo {volume} {4}},\
  \bibinfo {pages} {013194} (\bibinfo {year} {2022})}\BibitemShut {NoStop}%
\bibitem [{\citenamefont {Nakano}\ and\ \citenamefont
  {Adachi}(2024)}]{Nakano2024}%
  \BibitemOpen
  \bibfield  {author} {\bibinfo {author} {\bibfnamefont {H.}~\bibnamefont
  {Nakano}}\ and\ \bibinfo {author} {\bibfnamefont {K.}~\bibnamefont
  {Adachi}},\ }\bibfield  {title} {\bibinfo {title} {Universal properties of
  repulsive self-propelled particles and attractive driven particles},\ }\href
  {https://doi.org/10.1103/PhysRevResearch.6.013074} {\bibfield  {journal}
  {\bibinfo  {journal} {Phys. Rev. Res.}\ }\textbf {\bibinfo {volume} {6}},\
  \bibinfo {pages} {013074} (\bibinfo {year} {2024})}\BibitemShut {NoStop}%
\bibitem [{\citenamefont {Brambati}\ \emph {et~al.}(2022)\citenamefont
  {Brambati}, \citenamefont {Fava},\ and\ \citenamefont
  {Ginelli}}]{Brambati2022}%
  \BibitemOpen
  \bibfield  {author} {\bibinfo {author} {\bibfnamefont {M.}~\bibnamefont
  {Brambati}}, \bibinfo {author} {\bibfnamefont {G.}~\bibnamefont {Fava}},\
  and\ \bibinfo {author} {\bibfnamefont {F.}~\bibnamefont {Ginelli}},\
  }\bibfield  {title} {\bibinfo {title} {Signatures of directed and spontaneous
  flocking},\ }\href {https://doi.org/10.1103/PhysRevE.106.024608} {\bibfield
  {journal} {\bibinfo  {journal} {Phys Rev E}\ }\textbf {\bibinfo {volume}
  {106}},\ \bibinfo {pages} {024608} (\bibinfo {year} {2022})}\BibitemShut
  {NoStop}%
\bibitem [{\citenamefont {Solon}\ \emph {et~al.}(2022)\citenamefont {Solon},
  \citenamefont {Chat{\'e}}, \citenamefont {Toner},\ and\ \citenamefont
  {Tailleur}}]{Solon2022}%
  \BibitemOpen
  \bibfield  {author} {\bibinfo {author} {\bibfnamefont {A.}~\bibnamefont
  {Solon}}, \bibinfo {author} {\bibfnamefont {H.}~\bibnamefont {Chat{\'e}}},
  \bibinfo {author} {\bibfnamefont {J.}~\bibnamefont {Toner}},\ and\ \bibinfo
  {author} {\bibfnamefont {J.}~\bibnamefont {Tailleur}},\ }\bibfield  {title}
  {\bibinfo {title} {Susceptibility of polar flocks to spatial anisotropy},\
  }\href {https://doi.org/10.1103/PhysRevLett.128.208004} {\bibfield  {journal}
  {\bibinfo  {journal} {Phys. Rev. Lett.}\ }\textbf {\bibinfo {volume} {128}},\
  \bibinfo {pages} {208004} (\bibinfo {year} {2022})}\BibitemShut {NoStop}%
\bibitem [{\citenamefont {Chatterjee}\ \emph {et~al.}(2022)\citenamefont
  {Chatterjee}, \citenamefont {Mangeat},\ and\ \citenamefont
  {Rieger}}]{Chatterjee2022}%
  \BibitemOpen
  \bibfield  {author} {\bibinfo {author} {\bibfnamefont {S.}~\bibnamefont
  {Chatterjee}}, \bibinfo {author} {\bibfnamefont {M.}~\bibnamefont
  {Mangeat}},\ and\ \bibinfo {author} {\bibfnamefont {H.}~\bibnamefont
  {Rieger}},\ }\bibfield  {title} {\bibinfo {title} {Polar flocks with
  discretized directions: The active clock model approaching the vicsek
  model},\ }\href {https://doi.org/10.1209/0295-5075/ac6e4b} {\bibfield
  {journal} {\bibinfo  {journal} {EPL}\ }\textbf {\bibinfo {volume} {138}},\
  \bibinfo {pages} {41001} (\bibinfo {year} {2022})}\BibitemShut {NoStop}%
\bibitem [{\citenamefont {Mishra}\ \emph {et~al.}(2014)\citenamefont {Mishra},
  \citenamefont {Puri},\ and\ \citenamefont {Ramaswamy}}]{Mishra2014}%
  \BibitemOpen
  \bibfield  {author} {\bibinfo {author} {\bibfnamefont {S.}~\bibnamefont
  {Mishra}}, \bibinfo {author} {\bibfnamefont {S.}~\bibnamefont {Puri}},\ and\
  \bibinfo {author} {\bibfnamefont {S.}~\bibnamefont {Ramaswamy}},\ }\bibfield
  {title} {\bibinfo {title} {Aspects of the density field in an active
  nematic},\ }\href {https://doi.org/10.1098/rsta.2013.0364} {\bibfield
  {journal} {\bibinfo  {journal} {Philos. Trans. R. Soc. A}\ }\textbf {\bibinfo
  {volume} {372}},\ \bibinfo {pages} {20130364} (\bibinfo {year}
  {2014})}\BibitemShut {NoStop}%
\bibitem [{\citenamefont {Br{\"o}ker}\ \emph {et~al.}(2023)\citenamefont
  {Br{\"o}ker}, \citenamefont {Bickmann}, \citenamefont {Te~Vrugt},
  \citenamefont {Cates},\ and\ \citenamefont {Wittkowski}}]{Broker2023}%
  \BibitemOpen
  \bibfield  {author} {\bibinfo {author} {\bibfnamefont {S.}~\bibnamefont
  {Br{\"o}ker}}, \bibinfo {author} {\bibfnamefont {J.}~\bibnamefont
  {Bickmann}}, \bibinfo {author} {\bibfnamefont {M.}~\bibnamefont {Te~Vrugt}},
  \bibinfo {author} {\bibfnamefont {M.~E.}\ \bibnamefont {Cates}},\ and\
  \bibinfo {author} {\bibfnamefont {R.}~\bibnamefont {Wittkowski}},\ }\bibfield
   {title} {\bibinfo {title} {{Orientation-Dependent} propulsion of active
  brownian spheres: From {Self-Advection} to programmable cluster shapes},\
  }\href {https://doi.org/10.1103/PhysRevLett.131.168203} {\bibfield  {journal}
  {\bibinfo  {journal} {Phys. Rev. Lett.}\ }\textbf {\bibinfo {volume} {131}},\
  \bibinfo {pages} {168203} (\bibinfo {year} {2023})}\BibitemShut {NoStop}%
\bibitem [{\citenamefont {Othman}\ \emph {et~al.}()\citenamefont {Othman},
  \citenamefont {Midya}, \citenamefont {Auth},\ and\ \citenamefont
  {Gompper}}]{Othman2024}%
  \BibitemOpen
  \bibfield  {author} {\bibinfo {author} {\bibfnamefont {S.}~\bibnamefont
  {Othman}}, \bibinfo {author} {\bibfnamefont {J.}~\bibnamefont {Midya}},
  \bibinfo {author} {\bibfnamefont {T.}~\bibnamefont {Auth}},\ and\ \bibinfo
  {author} {\bibfnamefont {G.}~\bibnamefont {Gompper}},\ }\href
  {http://arxiv.org/abs/2403.02947} {\bibinfo {title} {Phase behavior and
  dynamics of active brownian particles in an alignment field}},\ \Eprint
  {https://arxiv.org/abs/2403.02947} {arXiv:2403.02947} \BibitemShut {NoStop}%
\bibitem [{\citenamefont {Doi}(1976)}]{Doi1976}%
  \BibitemOpen
  \bibfield  {author} {\bibinfo {author} {\bibfnamefont {M.}~\bibnamefont
  {Doi}},\ }\bibfield  {title} {\bibinfo {title} {Second quantization
  representation for classical many-particle system},\ }\href
  {https://doi.org/10.1088/0305-4470/9/9/008} {\bibfield  {journal} {\bibinfo
  {journal} {J. Phys. A}\ }\textbf {\bibinfo {volume} {9}},\ \bibinfo {pages}
  {1465} (\bibinfo {year} {1976})}\BibitemShut {NoStop}%
\bibitem [{\citenamefont {Peliti}(1985)}]{Peliti1985}%
  \BibitemOpen
  \bibfield  {author} {\bibinfo {author} {\bibfnamefont {L.}~\bibnamefont
  {Peliti}},\ }\bibfield  {title} {\bibinfo {title} {Path integral approach to
  birth-death processes on a lattice},\ }\href
  {https://doi.org/10.1051/jphys:019850046090146900} {\bibfield  {journal}
  {\bibinfo  {journal} {J. Phys. France}\ }\textbf {\bibinfo {volume} {46}},\
  \bibinfo {pages} {1469} (\bibinfo {year} {1985})}\BibitemShut {NoStop}%
\bibitem [{\citenamefont {T{\"a}uber}(2014)}]{Tauber2014}%
  \BibitemOpen
  \bibfield  {author} {\bibinfo {author} {\bibfnamefont {U.~C.}\ \bibnamefont
  {T{\"a}uber}},\ }\href@noop {} {\emph {\bibinfo {title} {Critical Dynamics: A
  Field Theory Approach to Equilibrium and {Non-Equilibrium} Scaling
  Behavior}}}\ (\bibinfo  {publisher} {Cambridge University Press},\ \bibinfo
  {year} {2014})\BibitemShut {NoStop}%
\bibitem [{\citenamefont {Gwa}\ and\ \citenamefont {Spohn}(1992)}]{Gwa1992}%
  \BibitemOpen
  \bibfield  {author} {\bibinfo {author} {\bibfnamefont {L.~H.}\ \bibnamefont
  {Gwa}}\ and\ \bibinfo {author} {\bibfnamefont {H.}~\bibnamefont {Spohn}},\
  }\bibfield  {title} {\bibinfo {title} {Six-vertex model, roughened surfaces,
  and an asymmetric spin hamiltonian},\ }\href
  {https://doi.org/10.1103/PhysRevLett.68.725} {\bibfield  {journal} {\bibinfo
  {journal} {Phys. Rev. Lett.}\ }\textbf {\bibinfo {volume} {68}},\ \bibinfo
  {pages} {725} (\bibinfo {year} {1992})}\BibitemShut {NoStop}%
\bibitem [{\citenamefont {Sandow}(1994)}]{Sandow1994}%
  \BibitemOpen
  \bibfield  {author} {\bibinfo {author} {\bibfnamefont {S.}~\bibnamefont
  {Sandow}},\ }\bibfield  {title} {\bibinfo {title} {Partially asymmetric
  exclusion process with open boundaries},\ }\href
  {https://doi.org/10.1103/physreve.50.2660} {\bibfield  {journal} {\bibinfo
  {journal} {Phys. Rev. E}\ }\textbf {\bibinfo {volume} {50}},\ \bibinfo
  {pages} {2660} (\bibinfo {year} {1994})}\BibitemShut {NoStop}%
\bibitem [{\citenamefont {Kim}(1995)}]{Kim1995}%
  \BibitemOpen
  \bibfield  {author} {\bibinfo {author} {\bibfnamefont {D.}~\bibnamefont
  {Kim}},\ }\bibfield  {title} {\bibinfo {title} {Bethe ansatz solution for
  crossover scaling functions of the asymmetric {XXZ} chain and the
  {Kardar-Parisi-Zhang-type} growth model},\ }\href
  {https://doi.org/10.1103/physreve.52.3512} {\bibfield  {journal} {\bibinfo
  {journal} {Phys. Rev. E}\ }\textbf {\bibinfo {volume} {52}},\ \bibinfo
  {pages} {3512} (\bibinfo {year} {1995})}\BibitemShut {NoStop}%
\bibitem [{\citenamefont {Essler}\ and\ \citenamefont
  {Rittenberg}(1996)}]{Essler1996}%
  \BibitemOpen
  \bibfield  {author} {\bibinfo {author} {\bibfnamefont {F.~H.~L.}\
  \bibnamefont {Essler}}\ and\ \bibinfo {author} {\bibfnamefont
  {V.}~\bibnamefont {Rittenberg}},\ }\bibfield  {title} {\bibinfo {title}
  {Representations of the quadratic algebra and partially asymmetric diffusion
  with open boundaries},\ }\href {https://doi.org/10.1088/0305-4470/29/13/013}
  {\bibfield  {journal} {\bibinfo  {journal} {J. Phys. A}\ }\textbf {\bibinfo
  {volume} {29}},\ \bibinfo {pages} {3375} (\bibinfo {year}
  {1996})}\BibitemShut {NoStop}%
\bibitem [{\citenamefont {Ashida}\ \emph {et~al.}(2020)\citenamefont {Ashida},
  \citenamefont {Gong},\ and\ \citenamefont {Ueda}}]{Ashida2020}%
  \BibitemOpen
  \bibfield  {author} {\bibinfo {author} {\bibfnamefont {Y.}~\bibnamefont
  {Ashida}}, \bibinfo {author} {\bibfnamefont {Z.}~\bibnamefont {Gong}},\ and\
  \bibinfo {author} {\bibfnamefont {M.}~\bibnamefont {Ueda}},\ }\bibfield
  {title} {\bibinfo {title} {{Non-Hermitian} physics},\ }\href
  {https://doi.org/10.1080/00018732.2021.1876991} {\bibfield  {journal}
  {\bibinfo  {journal} {Adv. Phys.}\ }\textbf {\bibinfo {volume} {69}},\
  \bibinfo {pages} {249} (\bibinfo {year} {2020})}\BibitemShut {NoStop}%
\bibitem [{\citenamefont {Murugan}\ and\ \citenamefont
  {Vaikuntanathan}(2017)}]{Murugan2017}%
  \BibitemOpen
  \bibfield  {author} {\bibinfo {author} {\bibfnamefont {A.}~\bibnamefont
  {Murugan}}\ and\ \bibinfo {author} {\bibfnamefont {S.}~\bibnamefont
  {Vaikuntanathan}},\ }\bibfield  {title} {\bibinfo {title} {Topologically
  protected modes in non-equilibrium stochastic systems},\ }\href
  {https://doi.org/10.1038/ncomms13881} {\bibfield  {journal} {\bibinfo
  {journal} {Nat. Commun.}\ }\textbf {\bibinfo {volume} {8}},\ \bibinfo {pages}
  {13881} (\bibinfo {year} {2017})}\BibitemShut {NoStop}%
\bibitem [{\citenamefont {Dasbiswas}\ \emph {et~al.}(2018)\citenamefont
  {Dasbiswas}, \citenamefont {Mandadapu},\ and\ \citenamefont
  {Vaikuntanathan}}]{Dasbiswas2018}%
  \BibitemOpen
  \bibfield  {author} {\bibinfo {author} {\bibfnamefont {K.}~\bibnamefont
  {Dasbiswas}}, \bibinfo {author} {\bibfnamefont {K.~K.}\ \bibnamefont
  {Mandadapu}},\ and\ \bibinfo {author} {\bibfnamefont {S.}~\bibnamefont
  {Vaikuntanathan}},\ }\bibfield  {title} {\bibinfo {title} {Topological
  localization in out-of-equilibrium dissipative systems},\ }\href
  {https://doi.org/10.1073/pnas.1721096115} {\bibfield  {journal} {\bibinfo
  {journal} {Proc. Natl. Acad. Sci. U.S.A.}\ }\textbf {\bibinfo {volume}
  {115}},\ \bibinfo {pages} {E9031} (\bibinfo {year} {2018})}\BibitemShut
  {NoStop}%
\bibitem [{\citenamefont {Thompson}\ \emph {et~al.}(2011)\citenamefont
  {Thompson}, \citenamefont {Tailleur}, \citenamefont {Cates},\ and\
  \citenamefont {Blythe}}]{Thompson2011}%
  \BibitemOpen
  \bibfield  {author} {\bibinfo {author} {\bibfnamefont {A.~G.}\ \bibnamefont
  {Thompson}}, \bibinfo {author} {\bibfnamefont {J.}~\bibnamefont {Tailleur}},
  \bibinfo {author} {\bibfnamefont {M.~E.}\ \bibnamefont {Cates}},\ and\
  \bibinfo {author} {\bibfnamefont {R.~A.}\ \bibnamefont {Blythe}},\ }\bibfield
   {title} {\bibinfo {title} {Lattice models of nonequilibrium bacterial
  dynamics},\ }\href {https://doi.org/10.1088/1742-5468/2011/02/P02029}
  {\bibfield  {journal} {\bibinfo  {journal} {J. Stat. Mech.}\ }\textbf
  {\bibinfo {volume} {2011}},\ \bibinfo {pages} {P02029} (\bibinfo {year}
  {2011})}\BibitemShut {NoStop}%
\bibitem [{\citenamefont {Peruani}\ \emph {et~al.}(2011)\citenamefont
  {Peruani}, \citenamefont {Klauss}, \citenamefont {Deutsch},\ and\
  \citenamefont {Voss-Boehme}}]{Peruani2011}%
  \BibitemOpen
  \bibfield  {author} {\bibinfo {author} {\bibfnamefont {F.}~\bibnamefont
  {Peruani}}, \bibinfo {author} {\bibfnamefont {T.}~\bibnamefont {Klauss}},
  \bibinfo {author} {\bibfnamefont {A.}~\bibnamefont {Deutsch}},\ and\ \bibinfo
  {author} {\bibfnamefont {A.}~\bibnamefont {Voss-Boehme}},\ }\bibfield
  {title} {\bibinfo {title} {Traffic jams, gliders, and bands in the quest for
  collective motion of self-propelled particles},\ }\href
  {https://doi.org/10.1103/PhysRevLett.106.128101} {\bibfield  {journal}
  {\bibinfo  {journal} {Phys. Rev. Lett.}\ }\textbf {\bibinfo {volume} {106}},\
  \bibinfo {pages} {128101} (\bibinfo {year} {2011})}\BibitemShut {NoStop}%
\bibitem [{\citenamefont {Whitelam}\ \emph {et~al.}(2018)\citenamefont
  {Whitelam}, \citenamefont {Klymko},\ and\ \citenamefont
  {Mandal}}]{Whitelam2018}%
  \BibitemOpen
  \bibfield  {author} {\bibinfo {author} {\bibfnamefont {S.}~\bibnamefont
  {Whitelam}}, \bibinfo {author} {\bibfnamefont {K.}~\bibnamefont {Klymko}},\
  and\ \bibinfo {author} {\bibfnamefont {D.}~\bibnamefont {Mandal}},\
  }\bibfield  {title} {\bibinfo {title} {Phase separation and large deviations
  of lattice active matter},\ }\href {https://doi.org/10.1063/1.5023403}
  {\bibfield  {journal} {\bibinfo  {journal} {J. Chem. Phys.}\ }\textbf
  {\bibinfo {volume} {148}},\ \bibinfo {pages} {154902} (\bibinfo {year}
  {2018})}\BibitemShut {NoStop}%
\bibitem [{\citenamefont {Kourbane-Houssene}\ \emph {et~al.}(2018)\citenamefont
  {Kourbane-Houssene}, \citenamefont {Erignoux}, \citenamefont {Bodineau},\
  and\ \citenamefont {Tailleur}}]{Kourbane-Houssene2018}%
  \BibitemOpen
  \bibfield  {author} {\bibinfo {author} {\bibfnamefont {M.}~\bibnamefont
  {Kourbane-Houssene}}, \bibinfo {author} {\bibfnamefont {C.}~\bibnamefont
  {Erignoux}}, \bibinfo {author} {\bibfnamefont {T.}~\bibnamefont {Bodineau}},\
  and\ \bibinfo {author} {\bibfnamefont {J.}~\bibnamefont {Tailleur}},\
  }\bibfield  {title} {\bibinfo {title} {Exact hydrodynamic description of
  active lattice gases},\ }\href
  {https://doi.org/10.1103/PhysRevLett.120.268003} {\bibfield  {journal}
  {\bibinfo  {journal} {Phys. Rev. Lett.}\ }\textbf {\bibinfo {volume} {120}},\
  \bibinfo {pages} {268003} (\bibinfo {year} {2018})}\BibitemShut {NoStop}%
\bibitem [{\citenamefont {Partridge}\ and\ \citenamefont
  {Lee}(2019)}]{Partridge2019}%
  \BibitemOpen
  \bibfield  {author} {\bibinfo {author} {\bibfnamefont {B.}~\bibnamefont
  {Partridge}}\ and\ \bibinfo {author} {\bibfnamefont {C.~F.}\ \bibnamefont
  {Lee}},\ }\bibfield  {title} {\bibinfo {title} {Critical {Motility-Induced}
  phase separation belongs to the ising universality class},\ }\href
  {https://doi.org/10.1103/PhysRevLett.123.068002} {\bibfield  {journal}
  {\bibinfo  {journal} {Phys. Rev. Lett.}\ }\textbf {\bibinfo {volume} {123}},\
  \bibinfo {pages} {068002} (\bibinfo {year} {2019})}\BibitemShut {NoStop}%
\bibitem [{\citenamefont {Mukherjee}\ \emph {et~al.}()\citenamefont
  {Mukherjee}, \citenamefont {Saha}, \citenamefont {Sadhu}, \citenamefont
  {Dhar},\ and\ \citenamefont {Sabhapandit}}]{Mukherjee2024}%
  \BibitemOpen
  \bibfield  {author} {\bibinfo {author} {\bibfnamefont {R.}~\bibnamefont
  {Mukherjee}}, \bibinfo {author} {\bibfnamefont {S.}~\bibnamefont {Saha}},
  \bibinfo {author} {\bibfnamefont {T.}~\bibnamefont {Sadhu}}, \bibinfo
  {author} {\bibfnamefont {A.}~\bibnamefont {Dhar}},\ and\ \bibinfo {author}
  {\bibfnamefont {S.}~\bibnamefont {Sabhapandit}},\ }\href
  {http://arxiv.org/abs/2405.19984} {\bibinfo {title} {Hydrodynamics of a
  hard-core non-polar active lattice gas}},\ \Eprint
  {https://arxiv.org/abs/2405.19984} {arXiv:2405.19984} \BibitemShut {NoStop}%
\bibitem [{Note1()}]{Note1}%
  \BibitemOpen
  \bibinfo {note} {Equation~\protect \eqref {eq_hydro} is a simpler version of
  Eq.~(2) in Ref.~\cite {Adachi2022} [or Eq.~(S-21) in Ref.~\cite
  {Mukherjee2024}], which is derived for a repulsive active lattice gas
  model.}\BibitemShut {Stop}%
\bibitem [{\citenamefont {Hohenberg}\ and\ \citenamefont
  {Halperin}(1977)}]{Hohenberg1977}%
  \BibitemOpen
  \bibfield  {author} {\bibinfo {author} {\bibfnamefont {P.~C.}\ \bibnamefont
  {Hohenberg}}\ and\ \bibinfo {author} {\bibfnamefont {B.~I.}\ \bibnamefont
  {Halperin}},\ }\bibfield  {title} {\bibinfo {title} {Theory of dynamic
  critical phenomena},\ }\href {https://doi.org/10.1103/RevModPhys.49.435}
  {\bibfield  {journal} {\bibinfo  {journal} {Rev. Mod. Phys.}\ }\textbf
  {\bibinfo {volume} {49}},\ \bibinfo {pages} {435} (\bibinfo {year}
  {1977})}\BibitemShut {NoStop}%
\bibitem [{\citenamefont {Chaikin}\ and\ \citenamefont
  {Lubensky}(1995)}]{Chaikin1995}%
  \BibitemOpen
  \bibfield  {author} {\bibinfo {author} {\bibfnamefont {P.~M.}\ \bibnamefont
  {Chaikin}}\ and\ \bibinfo {author} {\bibfnamefont {T.~C.}\ \bibnamefont
  {Lubensky}},\ }\href@noop {} {\emph {\bibinfo {title} {Principles of
  Condensed Matter Physics}}}\ (\bibinfo  {publisher} {Cambridge University
  Press},\ \bibinfo {year} {1995})\BibitemShut {NoStop}%
\bibitem [{\citenamefont {Speck}\ \emph {et~al.}(2015)\citenamefont {Speck},
  \citenamefont {Menzel}, \citenamefont {Bialk{\'e}},\ and\ \citenamefont
  {L{\"o}wen}}]{Speck2015}%
  \BibitemOpen
  \bibfield  {author} {\bibinfo {author} {\bibfnamefont {T.}~\bibnamefont
  {Speck}}, \bibinfo {author} {\bibfnamefont {A.~M.}\ \bibnamefont {Menzel}},
  \bibinfo {author} {\bibfnamefont {J.}~\bibnamefont {Bialk{\'e}}},\ and\
  \bibinfo {author} {\bibfnamefont {H.}~\bibnamefont {L{\"o}wen}},\ }\bibfield
  {title} {\bibinfo {title} {Dynamical mean-field theory and weakly non-linear
  analysis for the phase separation of active brownian particles},\ }\href
  {https://doi.org/10.1063/1.4922324} {\bibfield  {journal} {\bibinfo
  {journal} {J. Chem. Phys.}\ }\textbf {\bibinfo {volume} {142}},\ \bibinfo
  {pages} {224109} (\bibinfo {year} {2015})}\BibitemShut {NoStop}%
\bibitem [{\citenamefont {Sternheim}\ and\ \citenamefont
  {Walker}(1972)}]{Sternheim1972}%
  \BibitemOpen
  \bibfield  {author} {\bibinfo {author} {\bibfnamefont {M.~M.}\ \bibnamefont
  {Sternheim}}\ and\ \bibinfo {author} {\bibfnamefont {J.~F.}\ \bibnamefont
  {Walker}},\ }\bibfield  {title} {\bibinfo {title} {{Non-Hermitian}
  hamiltonians, decaying states, and perturbation theory},\ }\href
  {https://doi.org/10.1103/PhysRevC.6.114} {\bibfield  {journal} {\bibinfo
  {journal} {Phys. Rev. C}\ }\textbf {\bibinfo {volume} {6}},\ \bibinfo {pages}
  {114} (\bibinfo {year} {1972})}\BibitemShut {NoStop}%
\bibitem [{\citenamefont {Sasa}(2005)}]{Sasa2005}%
  \BibitemOpen
  \bibfield  {author} {\bibinfo {author} {\bibfnamefont {S.-I.}\ \bibnamefont
  {Sasa}},\ }\bibfield  {title} {\bibinfo {title} {Long range spatial
  correlation between two brownian particles under external driving},\ }\href
  {https://doi.org/10.1016/j.physd.2005.02.004} {\bibfield  {journal} {\bibinfo
   {journal} {Physica D}\ }\textbf {\bibinfo {volume} {205}},\ \bibinfo {pages}
  {233} (\bibinfo {year} {2005})}\BibitemShut {NoStop}%
\bibitem [{\citenamefont {Lefevere}\ and\ \citenamefont
  {Tasaki}(2005)}]{Lefevere2005}%
  \BibitemOpen
  \bibfield  {author} {\bibinfo {author} {\bibfnamefont {R.}~\bibnamefont
  {Lefevere}}\ and\ \bibinfo {author} {\bibfnamefont {H.}~\bibnamefont
  {Tasaki}},\ }\bibfield  {title} {\bibinfo {title} {High-temperature expansion
  for nonequilibrium steady states in driven lattice gases},\ }\href
  {https://doi.org/10.1103/PhysRevLett.94.200601} {\bibfield  {journal}
  {\bibinfo  {journal} {Phys. Rev. Lett.}\ }\textbf {\bibinfo {volume} {94}},\
  \bibinfo {pages} {200601} (\bibinfo {year} {2005})}\BibitemShut {NoStop}%
\bibitem [{\citenamefont {Luo}\ \emph {et~al.}(2023)\citenamefont {Luo},
  \citenamefont {Gu}, \citenamefont {Park}, \citenamefont {Fang}, \citenamefont
  {Kwon}, \citenamefont {Urue{\~n}a}, \citenamefont {Read~de Alaniz},
  \citenamefont {Helgeson}, \citenamefont {Marchetti},\ and\ \citenamefont
  {Valentine}}]{Luo2023}%
  \BibitemOpen
  \bibfield  {author} {\bibinfo {author} {\bibfnamefont {Y.}~\bibnamefont
  {Luo}}, \bibinfo {author} {\bibfnamefont {M.}~\bibnamefont {Gu}}, \bibinfo
  {author} {\bibfnamefont {M.}~\bibnamefont {Park}}, \bibinfo {author}
  {\bibfnamefont {X.}~\bibnamefont {Fang}}, \bibinfo {author} {\bibfnamefont
  {Y.}~\bibnamefont {Kwon}}, \bibinfo {author} {\bibfnamefont {J.~M.}\
  \bibnamefont {Urue{\~n}a}}, \bibinfo {author} {\bibfnamefont
  {J.}~\bibnamefont {Read~de Alaniz}}, \bibinfo {author} {\bibfnamefont
  {M.~E.}\ \bibnamefont {Helgeson}}, \bibinfo {author} {\bibfnamefont {C.~M.}\
  \bibnamefont {Marchetti}},\ and\ \bibinfo {author} {\bibfnamefont {M.~T.}\
  \bibnamefont {Valentine}},\ }\bibfield  {title} {\bibinfo {title}
  {Molecular-scale substrate anisotropy, crowding and division drive collective
  behaviours in cell monolayers},\ }\href
  {https://doi.org/10.1098/rsif.2023.0160} {\bibfield  {journal} {\bibinfo
  {journal} {J. R. Soc. Interface}\ }\textbf {\bibinfo {volume} {20}},\
  \bibinfo {pages} {20230160} (\bibinfo {year} {2023})}\BibitemShut {NoStop}%
\bibitem [{\citenamefont {Gu}\ \emph {et~al.}(2023)\citenamefont {Gu},
  \citenamefont {Fang},\ and\ \citenamefont {Luo}}]{Gu2023}%
  \BibitemOpen
  \bibfield  {author} {\bibinfo {author} {\bibfnamefont {M.}~\bibnamefont
  {Gu}}, \bibinfo {author} {\bibfnamefont {X.}~\bibnamefont {Fang}},\ and\
  \bibinfo {author} {\bibfnamefont {Y.}~\bibnamefont {Luo}},\ }\bibfield
  {title} {\bibinfo {title} {{Data-Driven} model construction for anisotropic
  dynamics of active matter},\ }\href
  {https://doi.org/10.1103/PRXLife.1.013009} {\bibfield  {journal} {\bibinfo
  {journal} {PRX Life}\ }\textbf {\bibinfo {volume} {1}},\ \bibinfo {pages}
  {013009} (\bibinfo {year} {2023})}\BibitemShut {NoStop}%
\bibitem [{\citenamefont {Skillin}\ \emph {et~al.}(2023)\citenamefont
  {Skillin}, \citenamefont {Kirkpatrick}, \citenamefont {Herbert},
  \citenamefont {Nelson}, \citenamefont {Hach}, \citenamefont {G{\"u}nay},
  \citenamefont {Khan}, \citenamefont {DelRio}, \citenamefont {White},\ and\
  \citenamefont {Anseth}}]{Skillin2023}%
  \BibitemOpen
  \bibfield  {author} {\bibinfo {author} {\bibfnamefont {N.~P.}\ \bibnamefont
  {Skillin}}, \bibinfo {author} {\bibfnamefont {B.~E.}\ \bibnamefont
  {Kirkpatrick}}, \bibinfo {author} {\bibfnamefont {K.~M.}\ \bibnamefont
  {Herbert}}, \bibinfo {author} {\bibfnamefont {B.~R.}\ \bibnamefont {Nelson}},
  \bibinfo {author} {\bibfnamefont {G.~K.}\ \bibnamefont {Hach}}, \bibinfo
  {author} {\bibfnamefont {K.~A.}\ \bibnamefont {G{\"u}nay}}, \bibinfo {author}
  {\bibfnamefont {R.~M.}\ \bibnamefont {Khan}}, \bibinfo {author}
  {\bibfnamefont {F.~W.}\ \bibnamefont {DelRio}}, \bibinfo {author}
  {\bibfnamefont {T.~J.}\ \bibnamefont {White}},\ and\ \bibinfo {author}
  {\bibfnamefont {K.~S.}\ \bibnamefont {Anseth}},\ }\bibfield  {title}
  {\bibinfo {title} {Stiffness anisotropy coordinates supracellular
  contractility driving long-range myotube-ecm alignment},\ }\bibfield
  {journal} {\bibinfo  {journal} {bioRxiv}\ }\href
  {https://doi.org/10.1101/2023.08.08.552197} {10.1101/2023.08.08.552197}
  (\bibinfo {year} {2023})\BibitemShut {NoStop}%
\end{thebibliography}
%

\end{document}